\documentclass{article}
\usepackage[]{graphicx}
\usepackage[]{color}
\usepackage{textcomp}
\usepackage{float}
\makeatletter
\def\maxwidth{ %
  \ifdim\Gin@nat@width>\linewidth
    \linewidth
  \else
    \Gin@nat@width
  \fi
}
\makeatother

\definecolor{fgcolor}{rgb}{0.345, 0.345, 0.345}
\newcommand{\hlkwb}[1]{\textcolor[rgb]{0.69,0.353,0.396}{#1}}%

\usepackage{framed}
\makeatletter
 {\par\unskip\endMakeFramed%
 \at@end@of@kframe}
\makeatother

\definecolor{shadecolor}{rgb}{.97, .97, .97}
\definecolor{messagecolor}{rgb}{0, 0, 0}
\definecolor{warningcolor}{rgb}{1, 0, 1}
\definecolor{errorcolor}{rgb}{1, 0, 0}
\newenvironment{knitrout}{}{} % an empty environment to be redefined in TeX

\usepackage{alltt}

\usepackage{subcaption}
\usepackage{hyperref}
\usepackage{caption}
\usepackage{cleveref}

\newcommand{\rinline}[1]{SOMETHING WRONG WITH knitr}
\IfFileExists{upquote.sty}{\usepackage{upquote}}{}
\begin{document}

\let\ref\autoref
\title{Improving axial resolution in SIM using deep learning}
\author{Boland, Miguel A.; Cohen, Edward A. K.; Flaxman, Seth R.; Neil, Mark A. A.}

\maketitle
\begin{abstract}
Structured Illumination Microscopy is a widespread methodology to image live and fixed biological structures smaller than the diffraction limits of conventional optical microscopy. Using recent advances in image up-scaling through deep learning models, we demonstrate a method to reconstruct 3D SIM image stacks with twice the axial resolution attainable through conventional SIM reconstructions. We further demonstrate our method is robust to noise \& evaluate it against two point cases and axial gratings. Finally, we discuss potential adaptions of the method to further improve resolution.
\end{abstract}

\maketitle

\section{Introduction}
Structured Illumination Microscopy (SIM) is a super-resolution technique which enables a two-fold increase in lateral resolution (X/Y axis, perpendicular to the line of sight of the microscope) when compared to conventional fluorescence microscopy  \cite{heintzmann_1999} \cite{gustafsson_sim}. SIM functions via the illumination of structured light onto a specimen to obtain spectral information that would be out of the range visible to a wide-field microscope. Repeating this procedure across a set of structured light patterns constructs  a stack of raw SIM images of a specimen, the ensemble of which contains a wider spectral range of frequencies than obtained via a single exposure. Image stacks are then reconstructed into a single super-resolved image using computational algorithms. A wide range of illumination patterns can be used, including 3D patterns that allow for improvements in axial resolution (Z axis, parallel to line of sight of the microscope) \cite{heintzmann_2018}. This has proven to be a useful investigative tool for the analysis of live biological processes as it avoids most sample damage traditionally associated with electron microscopy \cite{Fiolka5311}. Example applications include in-vivo imaging of synapses in mice \cite{Turcotte9586}, or analysing the distribution of proteins on chromosomes, a task previously unfeasible due to the relative scale of the protein and a confocal microscope's diffraction limit \cite{Wang905}. As such, the further improvement of image reconstruction methodologies for microscopy may allow the observation of previously unobserved live biological processes. \newline

Improvements in axial resolution have historically been achieved by advances in optical setup and illumination techniques, such as 3D-SIM \cite{Gustafsson2008}, TIRF-SIM \cite{tirf_sim}, spot-scanning SR-SIM \cite{spot_scan_sr_sim} or 2-photon SR-SIM \cite{2phot_sr_sim}. As these techniques can incur significant costs and complexity, recent developments in deep learning have been adapted and optimised to address these issues from a purely computational approach. Proposed deep learning methods for SIM have successfully demonstrated reductions in aberrations and reconstruction artifacts due to movements of the specimen \cite{Forster:16} \cite{Forster:18} or low-light conditions \cite{christensen2020mlsim} \cite{unet_sim} \cite{cyclegan_sim}, but as of yet, deep learning networks have not been applied to the problem of reconstructing SIM image stacks to a resolution beyond that attainable by standard SIM methodologies. Training a network to take as its input a set of SIM images as might be produced by a 2D-SIM microscope setup (2-beam, 3-angle) and with a target that has the resolution characteristics of a 3D-SIM microscope, would allow the network to achieve an axial resolution comparable to 3D SIM microscopes without the motion artifacts or photo-bleaching caused by high image counts \cite{Fan2019}. \newline

As the range of imaged light frequencies is reduced in this microscope setup, the use of convolutional layers will allow us to infer additional information (as is currently done through Richardson-Lucy deconvolution \cite{Chakrova}). Building on the work of deep learning networks applied to SIM reconstructions, a modified Residual Channel Attention Network (RCAN) \cite{rcan_zhang} deep neural net is trained to reconstruct SIM image stacks at double the axial resolution than that attainable via the original SIM reconstruction methodology \cite{gustafsson_sim}. High resolution images are simulated as confocal images with an illumination wavelength reduced by a factor of $\sqrt{2}$ to match the lateral resolution obtained via seminal SIM reconstructions, whilst axial resolution is improved by decreasing the axial slice thickness of each image in the stack. Simulated training and test data is generated from both dense and biologically realistic 3D point clouds, at both standard resolution (SR) and high resolution (HR) as detailed in \Cref{sec:methods}. The network is evaluated against test data to investigate the method's effectiveness at resolving structures composed of point illuminations (\Cref{sec:results}). As we expect the network to output spatial frequencies which are not inherent in the input data (which is already achieved through deconvolution techniques \cite{Chakrova}), the network should be able to infer these from the widefield images using the mapping from simulated widefield images to simulated high resolution confocal images. 
Further analysis is performed that demonstrates robustness to low-lighting conditions (\Cref{sec:results}). Limitations in network generalisability and the broader potential of the work is discussed in \Cref{sec:discussion}.

\section{Existing work}
Computational methods for SIM image reconstruction have set benchmarks of attainable resolution, and demonstrated flexibility to imprecise or varying illumination patterns. Examples of these are blind-SIM \cite{jost}, in which gradient descent is used to minimise the least squared error between the measured data and data predicted by the convolution of an estimated biological sample with an estimated illumination pattern. By alternatively fixing the estimated data and estimated illumination pattern, a high resolution image can be retrieved with limited knowledge of the illumination pattern but at the cost of significant compute time. Similarly, the use of Richardson-Lucy deconvolutions \cite{Chakrova} to further improve the quality of the seminal SIM algorithm \cite{gustafsson_sim} have demonstrated significant improvements in resolution when experimentally measuring the PSF of the optical apparatus.

In contrast, deep learning has been applied to SIM reconstruction with varying goals, such as accelerating the imaging process by reducing the required number of raw frames (i.e the number of variations in illumination patterns) and inferring the spectral information from a smaller subset of SIM frames. These demonstrate an objective of high-throughput imaging, where a known microscope configuration is coupled with deep learning methods to minimise the imaging and processing time whilst improving the obtained resolution.

% Blind-SIM https://journals.plos.org/plosone/article/file?id=10.1371/journal.pone.0132174&type=printable
%     - DL is parameter free, can be processed on GPUs for high-throughput operation
% Richardson-Lucy
%     - No parameters required but multiple itera

This has been achieved using U-Nets \cite{unet_sim} and CycleGAN (a variation of a Conditional Generative Adversarial Network) \cite{cyclegan_sim}. These diverge from our goal which is to attain a higher axial resolution whilst maintaining a $2\times$ increase in lateral resolution.
Image super-resolution for generic photography has progressed considerably in the past decade, resulting in a large variety of architectures and techniques \cite{Wang_2020_dl_survey}. Notably, these form paradigms regarding the pre or post upsampling of inputs (relative to convolutional or other image processing), methods of image interpolation, residual learning, network recursiveness or parallel paths in addition to a large number of architecture-specific features. 

Amongst these exists the Residual Channel Attention Network (RCAN) \cite{rcan_zhang}, which is engineered to use inter-channel features (RGB colors in the original network) in addition to sub-pixel upsampling. These are particularly suited to SIM reconstruction, as this process entails the recombination of multiple raw images (i.e channels) into an up-scaled image. An RCAN for SIM reconstruction was successfully trained using simulated SIM stacks of generic stock imagery \cite{christensen2020mlsim}, which forms the basis for our architecture.

\section{Methods}
\label{sec:methods}
\subsection{Deep learning}

\subsubsection{Network architecture}
\label{subsection:network_architecture}

\begin{figure}
\includegraphics[width=5in]{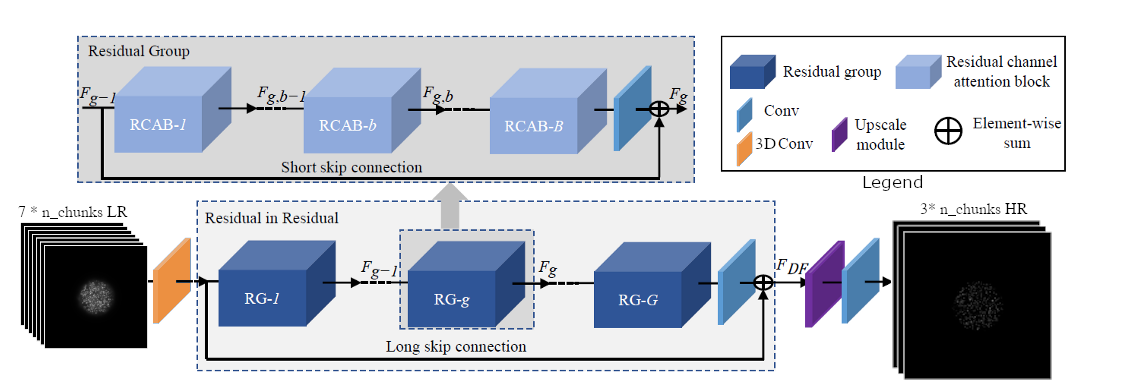}
\caption{Modified RCAN network architecture, adapted from \cite{rcan_zhang}. Note the use of a 3D convolution at the input, and the use of a single input channel and 3 output channels.}
\label{fig:rcan}
\end{figure}
The network architecture is based on the RCAN  \cite{rcan_zhang} which was previously used to implement a form of SIM reconstruction without up-scaling \cite{christensen2020mlsim}. We modified the network to re-implement the lateral up-scaling module found in the original implementation, therefore doubling the lateral output image size. The up-scaling module was necessary to generate images comparable to our target data, which is Nyquist sampled without aliasing. The use of an up or down-scaling algorithm to adapt the target to the network output or vice-versa could cause undesirable patterns to be learned by the network, whilst using the up-scaling module between learned layers allows for some learned adaption to the underlying mapping of the low-resolution image stacks and high-resolution confocal target images.  

Furthermore, the first 2D convolutional layer was replaced by a 3D convolution in order to infer information from other SIM stacks within the same processing chunk (see section below for chunk-generating procedure) prior to higher order operations (which take place in the residual blocks of the network). The number of output channels was also changed according to the chunk size, such that 3 consecutive gray scale (i.e.~single channel) images were produced for every inputted SIM stack.

The final network configuration utilised 12 groups of 3 residual blocks, a kernel size of 3x3 and a chunk size of 3; these parameters were chosen experimentally as detailed in Supplementary Materials - 6 - Network architecture \& Optimisation.

\subsubsection{Chunk processing}
Input SIM stacks were processed in chunks to allow varying depths of image stacks to be compatible with the network and improve the model's performance by inferring relations between consecutive axial images. In this scheme, a chunk size of two corresponds to a network input of 14 images (7 frames per SIM image * 2 chunk size) and a network output of 6 frames (3 frames per SIM image * 2 chunk size). Chunk size was determined experimentally by modifying the training data and input channels for the architecture described in  \Cref{sec:methods}; these were then evaluated using 24 withheld chromatin structure images and a series of images of spherical point clouds with increasing point densities. The evaluation of chunk size on network performance can be found in the supplementary materials Section A5.

% TODO re-write this section
The chunk size had no significant effect on the network's Mean Squared Error (MSE) (Mann-Whitney U-Test, p>0.05 for all pairs of chunked data), but the Structural Similarity Index (SSIM, a comparative metric of conservation of structural information between an image and reference image \cite{ssim}) was higher for a chunk size of 3 (U-Test, p<0.05 for 6/9 comparisons). This indicates that a chunk size of 3 maximises the conservation of the structure of the images, as reflected by the mean lateral FWHM (smaller than all other chunk size's means, U-Test, p<0.05) and axial mean (smaller than all other chunk sizes, U-Test, p<0.05).

\subsubsection{Data pre \& post processing}
Pairs of SIM image stacks and HR confocal images were axially sliced to produce chunked sets of images as described above. As the simulated images extended beyond the simulated underlying data, image pairs with little signal (mean pixel value below \(5\times10^{-7}\)) were discarded. The remaining set of chunked images was then normalised to a range of \emph{[0-1]}. 

A total of 100 chromatin structures and 68 spherical datasets were simulated to generate 168 image stacks. Of these, 10 chromatin structures and 20 spherical datasets were reserved for a test dataset, leaving 138 stacks for training \& validation. The input/output images were then axially chunked into sets of 3 SIM images, generating 12 training datapoints per stack for a total of 1656 training points. As the simulated stacks contain little signal at the start and end of the image stacks, chunks with a mean pixel value (measured as a 16-bit float) below $10^{-7}$ are discarded, leaving a total of 1178 training points and 130 validation points.

Negative pixel values did occur in the RCAN outputs due to small normally distributed random errors in output values; as a majority of test images are zero-valued backgrounds, many of the errors cause small negative values. The output was therefore post-processed to truncate negative values, which is in effect the equivalent to using an additional \emph{reLu} layer on the network but without causing a loss of learning gradient. 

\subsubsection{Training methodology}
All deep learning models were implemented using the PyTorch library (version 1.5) \cite{paszke2019pytorch} in Python 3.7. The starting learning rate was set to $10^{-6}$, was adjusted if the validation rate plateaued during training through the use of the \emph{ReduceLROnPlateau} method provided by PyTorch, and a further 1\% decrease was applied every 5 epochs to ensure fine tuning in later epochs. Network gradients were clipped at $\pm$ 0.1 to avoid exploding gradients. Training / validation sets were drawn from all generated data except for withheld test cases, which were randomly selected prior to training. Training took approximately 8 hours on a set of 2 Nvidia GTX 2080 TI GPUs. Over-fitting does not appear to have occurred in our training scheme as demonstrated by the continual decrease in mean MSE values for the validation dataset (see Supplementary Materials A6).

\subsection{Data simulation}
\label{subsec:data_simulation}
The model was trained using simulated SIM image stacks as input, and an up-sampled confocal image as ground truth. The methods used to generate these and benchmark SIM reconstructions are described below.

\subsubsection{SIM image stacks}
Raw SIM images were generated from 3D point cloud data; these simulated a biologically realistic chromatin structure \cite{wang_chromatin} (approximately bounded in a square of $16\times16$ \textmu m) or a point cloud encompassed in a sphere (radius of 5 \textmu m) with varying numbers of point emitters. The simulated microscope utilised a 7-frame setup based on the interference of 3 coherent beams to produce a 2D hexagonal illumination pattern on the sample; this is well illustrated in Figure 2 D-F of Ingerman et. al., 2019 \cite{heintzmann_2018}. The structured illumination is simulated as a thin light-sheet projected onto the specimen from the side \cite{sala_20} thereby avoiding the issues of reduced resolution enhancement usually found with this SIM geometry in an epi-illumination setup. The simulated microscope setup and image simulation approach is further described by Gong et al \cite{gong}, and a mathematical description can be found in the supplementary material of Gong et al. The system was simulated with a camera size of 256x256 pixels of size 6.5 $\mu$m square, an objective magnification of 60, with a numerical aperture of 1.1 and an illumination/emission wavelength of 525 nm. The chosen illumination/emission wavelengths are not required to be identical and were chosen for convenience in this experiment. No averse effects are expected from the use of wavelengths differing by up to 10\%. Photobleaching and other image quality degradation were not simulated as part of the experiment, though low-lighting conditions were simulated using Poisson statistics.

\subsubsection{High-resolution confocal microscope simulation}
Target output stacks were generated using a simulation of a confocal microscope on the same cloud point data. The parameters of a simulated confocal microscope were set to out-perform the simulated SIM microscope (halved pixel size, illumination wavelength is divided by \(\sqrt{2}\), axial \& lateral Nyquist sampling rate doubled, squared PSF function). These parameters ensure a doubling in both axial and lateral resolutions.

The number of output frames was also increased from 40 to 120 frames in order to improve the precision of measurements of axial resolution.

\subsubsection{SIM reconstructions}
 While complete high resolution images can be reconstructed with a set of just 7 images, this hexagonal patterned SIM illumination reconstructs according to the same carrier spatial frequencies as are found in a 2-beam, 3-angle setup using 9-frames  \cite{heintzmann_1999} \cite{gustafsson_sim}.  As a consequence the processing of the 7-frame data to a super-resolved image with enhanced lateral resolution \cite{gong} was achieved in a similar way to the standard SIM reconstruction algorithm \cite{gustafsson_sim}. The estimation of reconstruction parameters was found to be imprecise due to the sub-sampling of image stacks in which the number of point emitters was limited. Constant reconstruction parameters matching those used to simulate the images were therefore used to reduce any source of variability in the reconstructed SIM images. Reconstructed SIM images were calculated to provide a comparative baseline for our RCAN-reconstructed images. This was implemented in Python (see \Cref{sec:data_access}), and was modified to output 120 frames to match the number of output frames in our high resolution confocal simulations and network outputs.

\subsection{Measuring resolution}
The Full Width at Half Maximum (FWHM) metric has commonly been used to evaluate the resolution of microscopy images \cite{Gustafsson2008} \cite{Wegel2016}. It is defined as the width of a point spread function in a given direction at half the peak intensity; a large FWHM would indicate a wider point spread function (PSF) resulting in a less detailed image, whilst a smaller FWHM reflects a narrower PSF and a sharper image. Alternative methods such as fourier-ring correlation \cite{Koho535583} could be viable to measure resolution; we instead elected to use FWHM as it provided a high-throughput measure of resolution for images of point clouds. \newline

The FWHM was estimated using a method based on auto-correlation functions \cite{autocorrelation_func}, thereby evaluating the FWHM for every point light source in an image rather than individually and manually selected data points. The auto-correlation function (ACF) of an image was calculated using the inverse Fourier transform of the power spectrum on individual 3D datasets. High densities of data-points can cause significant cross-correlation of signal from neighbouring PSFs; these are removed by modelling the image auto-correlation as the sum of two Gaussians. The first Gaussian models the auto-correlation of individual PSFs in the image, whilst the second models the cross-correlation between neighbouring randomly distributed PSFs. These are estimated using a bounded Least-Squares regression, and the FWHM is calculated from the Gaussian with the smallest standard deviation. An example of the modelling process is illustrated in supplementary materials Section A4. The Python \& Scipy implementation of this methodology is available via GitHub, and an equivalent mathematical notation is available in the supplementary materials Section A1. 

\section{Results}
\label{sec:results}
\subsection{Doubling axial resolution using RCAN networks}
\label{subsec:axial_resolution}

\begin{figure}[H]
\centering
\begin{subfigure}{.45\textwidth}
\begin{knitrout}
\definecolor{shadecolor}{rgb}{0.969, 0.969, 0.969}\color{fgcolor}
\includegraphics[width=\maxwidth]{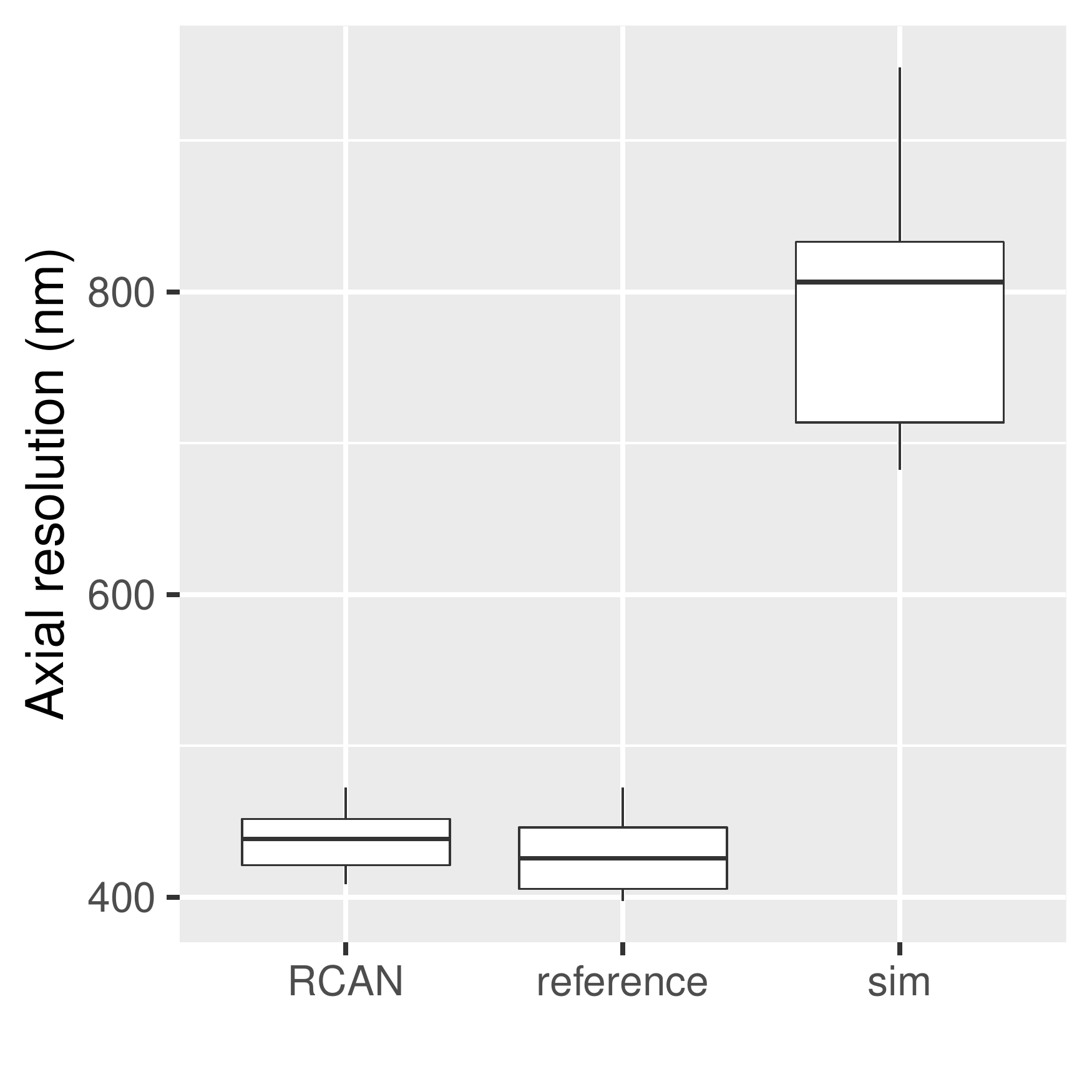} 

\end{knitrout}
    \caption{}
    \label{fig:axial_doubling_struct}
\end{subfigure}
\begin{subfigure}{.45\textwidth}
\begin{knitrout}
\definecolor{shadecolor}{rgb}{0.969, 0.969, 0.969}\color{fgcolor}
\includegraphics[width=\maxwidth]{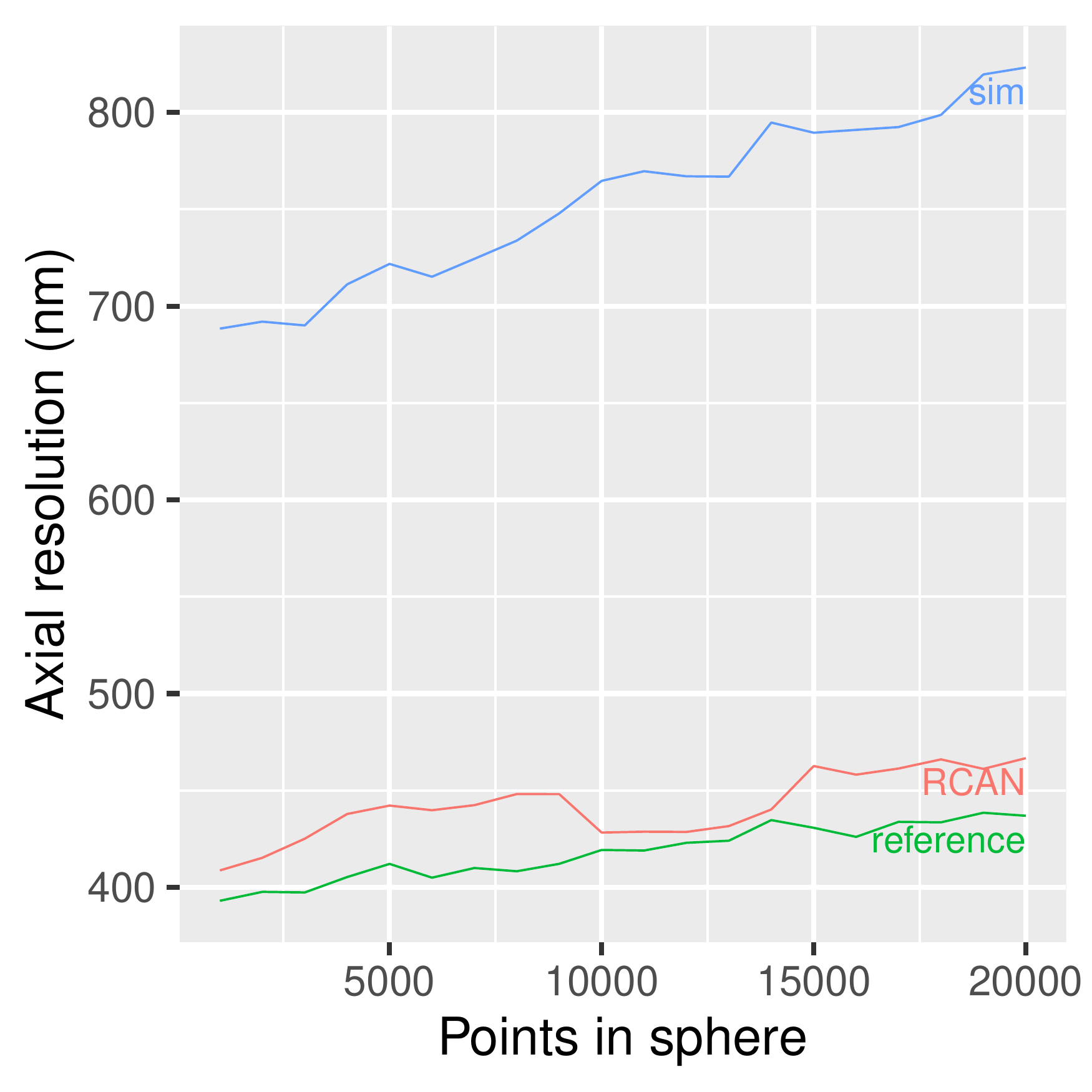} 

\end{knitrout}
    \caption{}
    \label{fig:axial_doubling_points}
\end{subfigure}
\begin{subfigure}{.3\textwidth}
\begin{knitrout}
\definecolor{shadecolor}{rgb}{0.969, 0.969, 0.969}\color{fgcolor}
\includegraphics[width=\maxwidth]{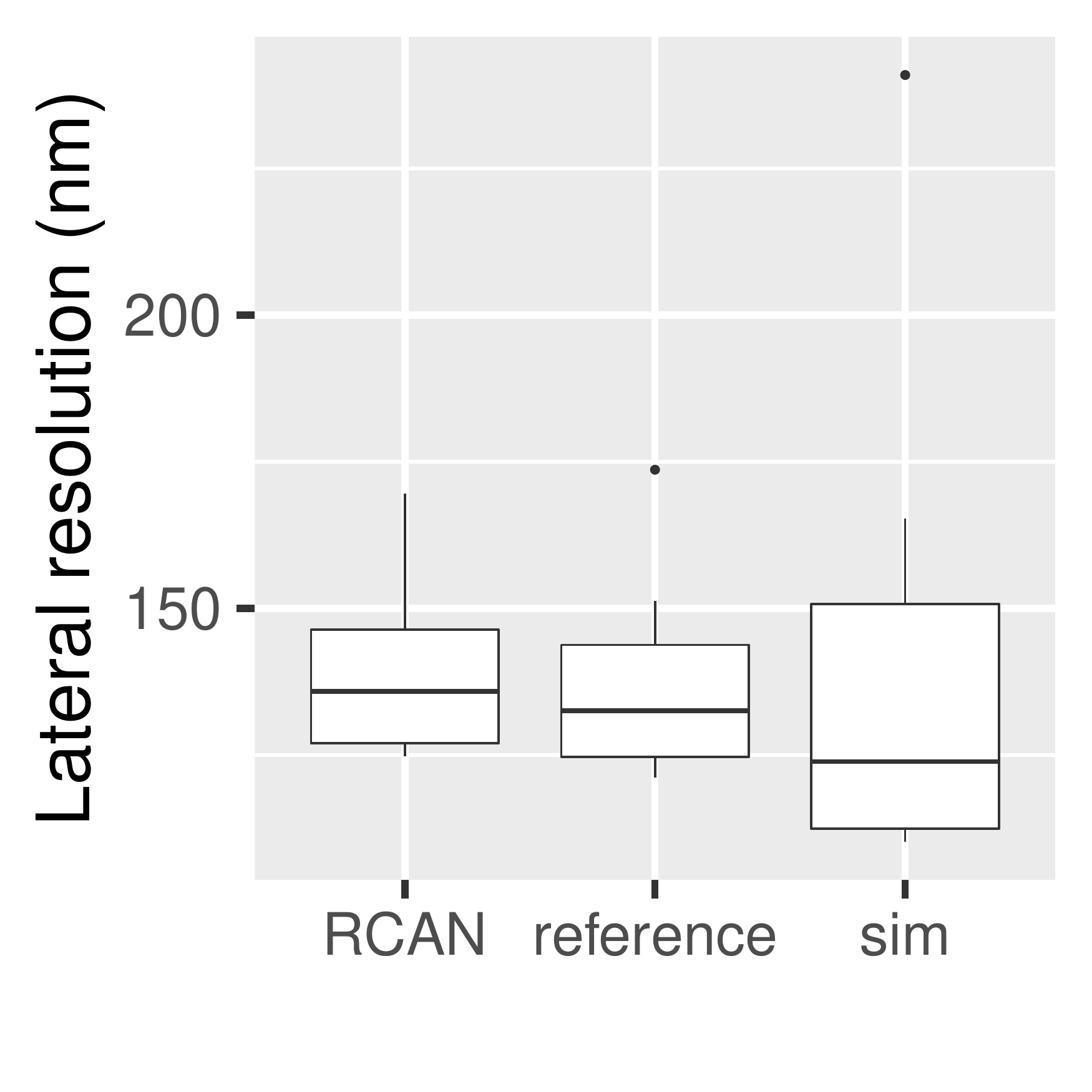} 

\end{knitrout}
    \caption{}
    \label{fig:axial_doubling_lateral}
\end{subfigure}
\begin{subfigure}{.3\textwidth}
\begin{knitrout}
\definecolor{shadecolor}{rgb}{0.969, 0.969, 0.969}\color{fgcolor}
\includegraphics[width=\maxwidth]{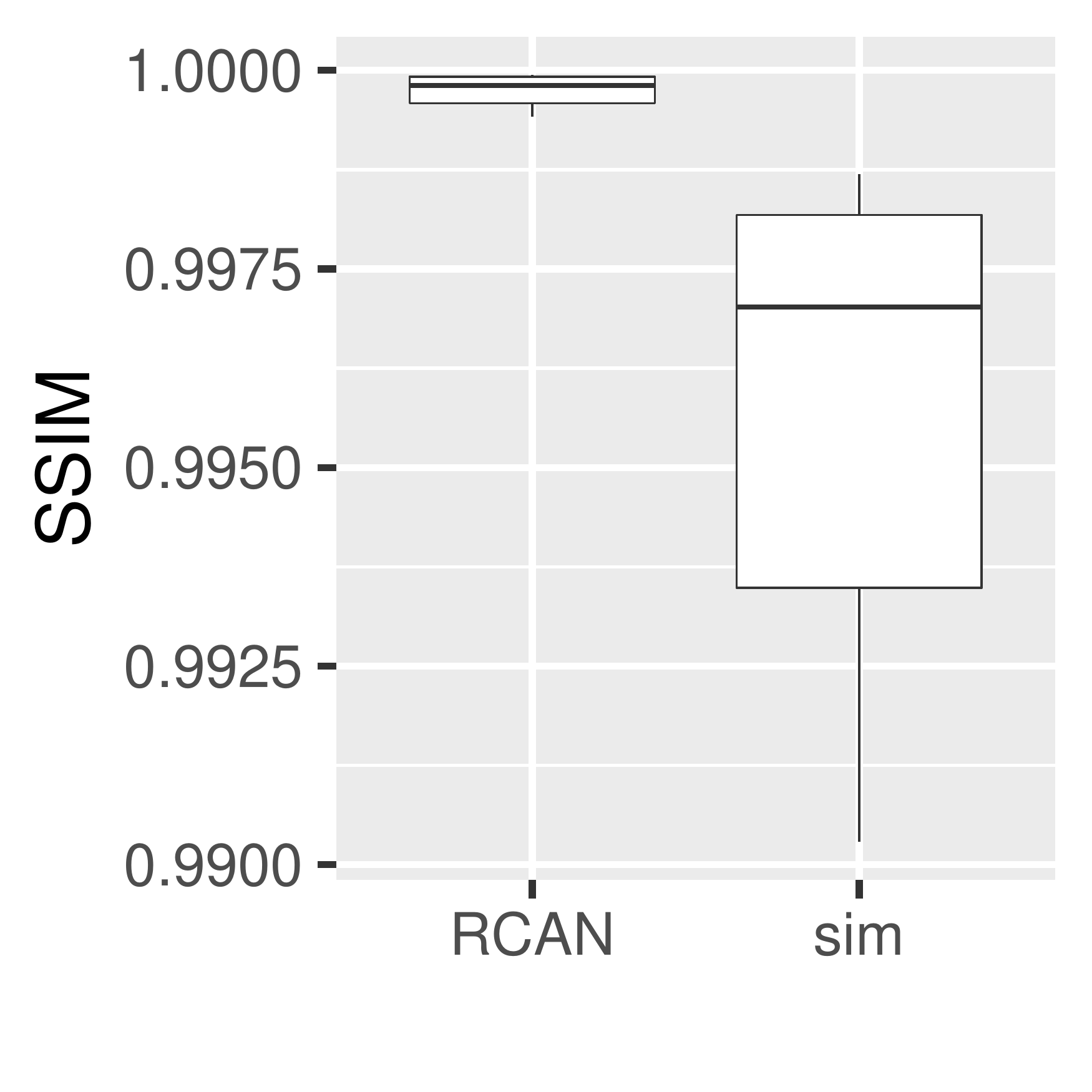} 

\end{knitrout}
    \caption{}
    \label{fig:axial_doubling_ssim}
\end{subfigure}
\begin{subfigure}{.3\textwidth}
\begin{knitrout}
\definecolor{shadecolor}{rgb}{0.969, 0.969, 0.969}\color{fgcolor}
\includegraphics[width=\maxwidth]{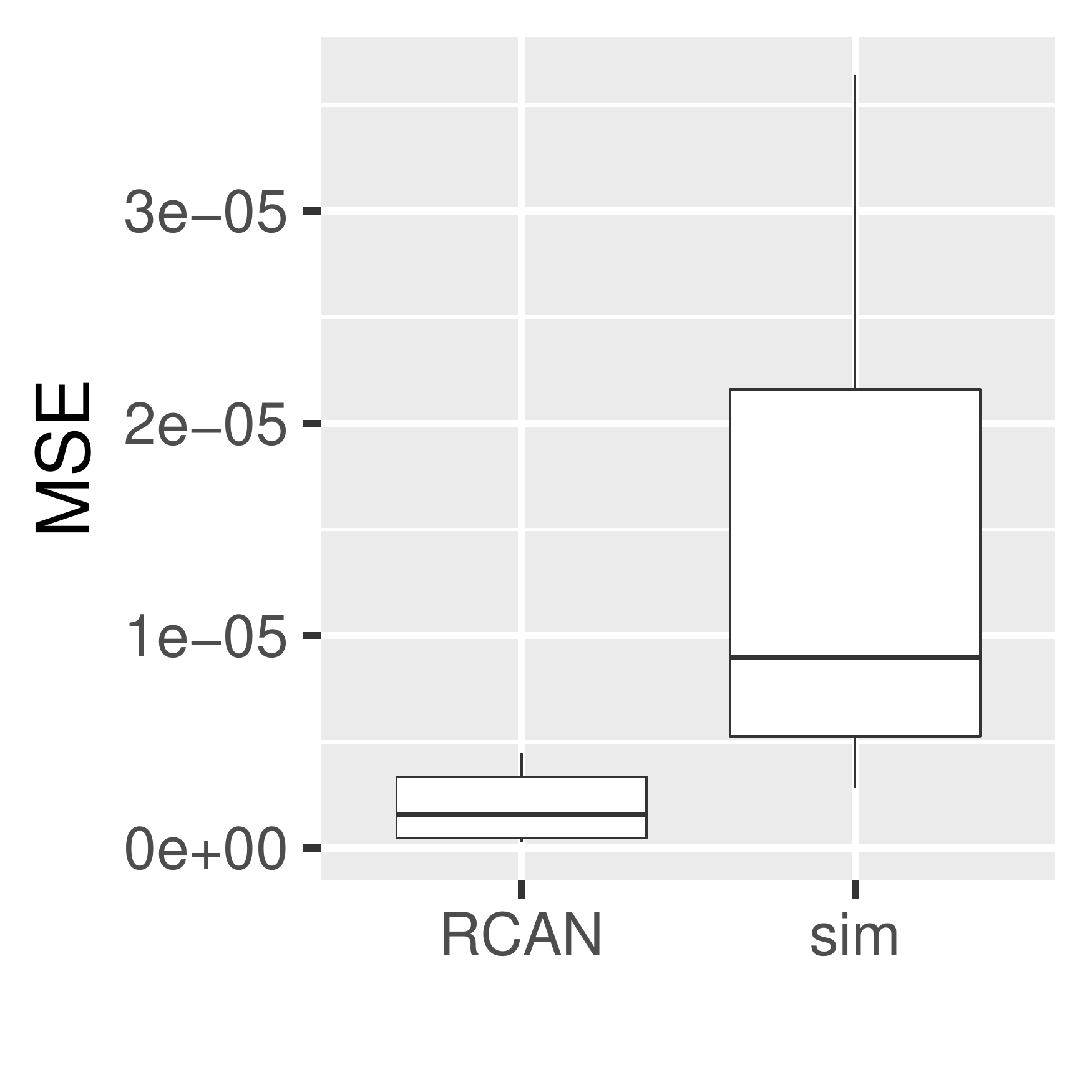} 

\end{knitrout}
    \caption{}
    \label{fig:axial_doubling_mse}
\end{subfigure}
\caption{
The proposed deep learning model (see \Cref{sec:methods}) demonstrates a doubled axial resolution compared to a standard SIM reconstruction when evaluated against 24 simulated chromatin structures (a) (whiskers represent percentiles, boxes represent IQR). An equivalent improvement in resolution is observed when reconstructing images of point clouds of increasing densities (b). Lateral resolution is maintained when compared to SIM reconstructions (c). Finally the RCAN  (mean=0.999, std=0.0002) appears to better preserve the structure of the underlying data compared to SIM (mean=0.995, std=0.003) as measured by SSIM \cite{ssim} (T-test t(10)=4.05, p<0.05), and MSE was marginally reduced (RCAN (mean=1.96e-06, std=1.69e-06), SIM (mean=1.43e-05, std=1.18e-05), T-test t(10)=-3.2657, p<0.01).}
\label{fig:axial_doubling}
\end{figure}

The network's ability to double the axial resolution was tested using 24 unseen simulated chromatin structures; this is visible in \ref{fig:axial_doubling_struct}. The mean axial FWHM for RCAN reconstructions (mean=436.8nm, std=21.8) was found to be significantly smaller than those produced using SIM (mean=669.67nm, std=207.2) (T-Test t(10)=-3.5, p<0.05). We further found the model to be robust to increasing point density when compared to standard SIM reconstructions (see \ref{fig:axial_doubling_points}). This is achieved without significantly affecting the lateral resolution (X \& Y axis, perpendicular to line of sight of the microscope), as seen in \ref{fig:axial_doubling_lateral}, and maintaining a strong conservation of structure as seen in \ref{fig:axial_doubling_ssim}. This is illustrated by the orthogonal Y/Z view in the supplementary material Section A2 - Figure 2, where the high structural fidelity and low background noise of the RCAN output contrasts with the SIM reconstruction. \newline

\subsection{Robustness in low-light imaging conditions}
\label{subsec:noise_robustness}

\begin{figure}[H]
\centering
\begin{subfigure}{.45\textwidth}
\begin{knitrout}
\definecolor{shadecolor}{rgb}{0.969, 0.969, 0.969}\color{fgcolor}
\includegraphics[width=\maxwidth]{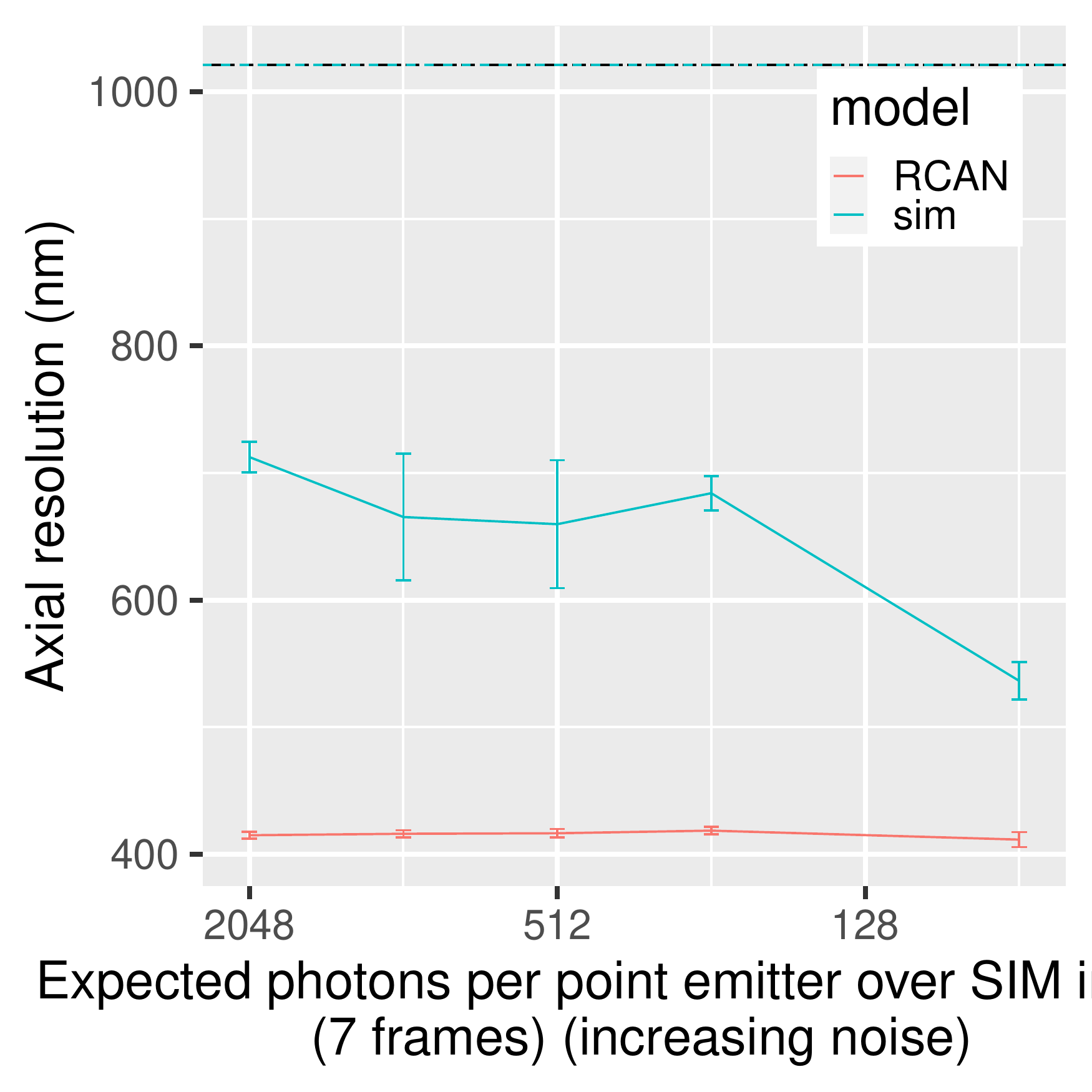} 

\end{knitrout}
    \caption{}
    \label{fig:noise_z}
\end{subfigure}
\begin{subfigure}{.45\textwidth}
\begin{knitrout}
\definecolor{shadecolor}{rgb}{0.969, 0.969, 0.969}\color{fgcolor}
\includegraphics[width=\maxwidth]{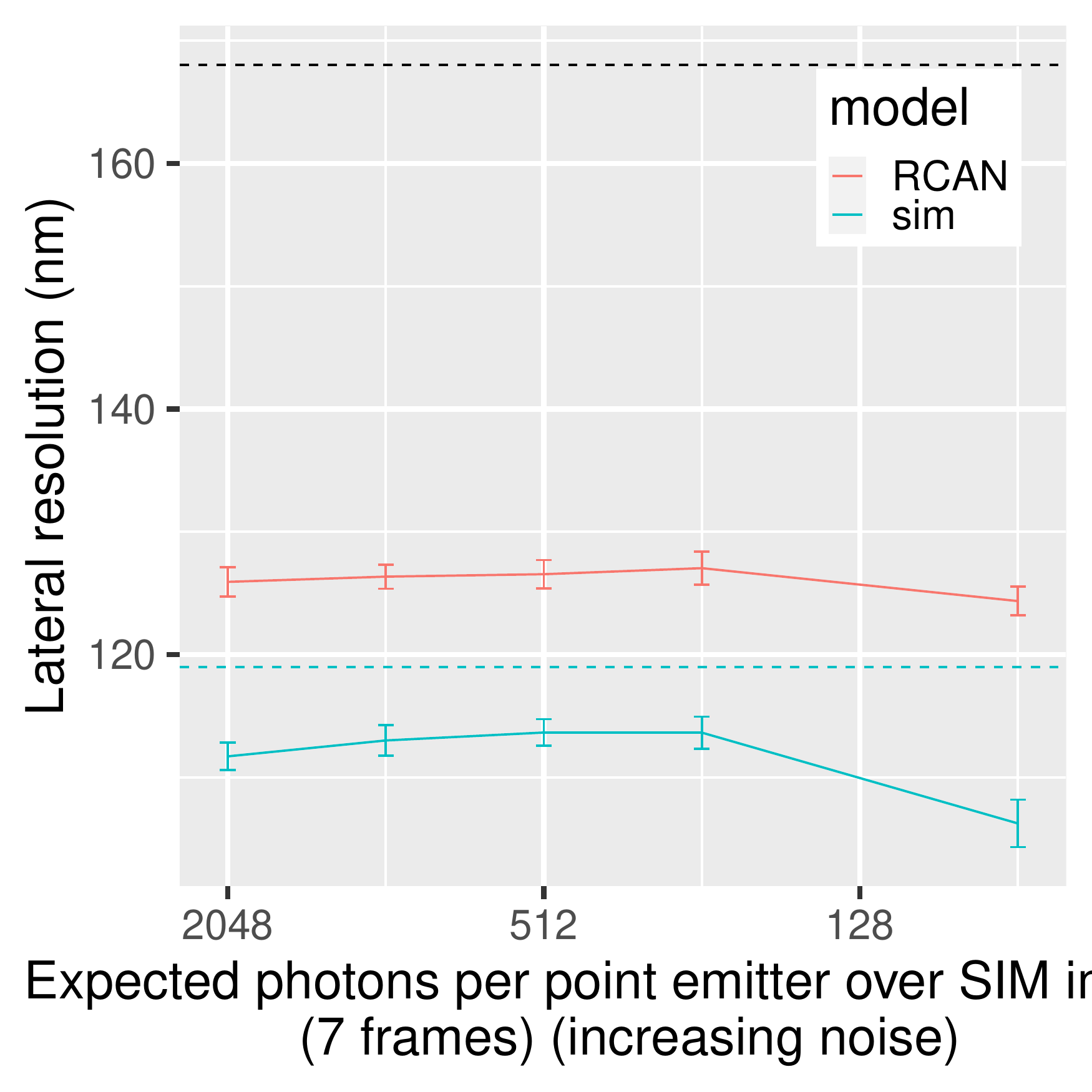} 

\end{knitrout}
    \caption{}
    \label{fig:noise_xy}
\end{subfigure}
\begin{subfigure}{.45\textwidth}
\begin{knitrout}
\definecolor{shadecolor}{rgb}{0.969, 0.969, 0.969}\color{fgcolor}
\includegraphics[width=\maxwidth]{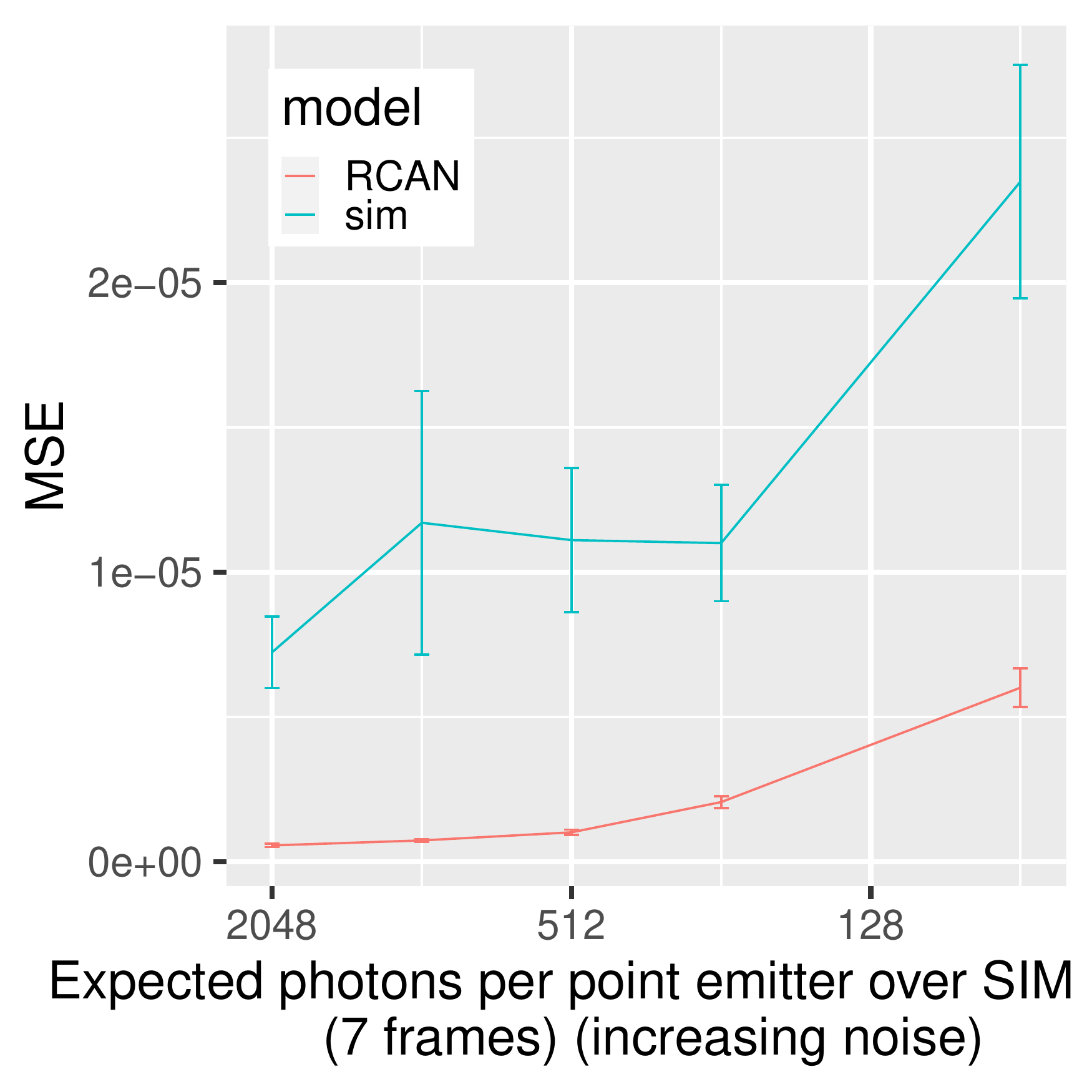} 

\end{knitrout}
    \caption{}
    \label{fig:noise_mse}
\end{subfigure}
\begin{subfigure}{.45\textwidth}
\begin{knitrout}
\definecolor{shadecolor}{rgb}{0.969, 0.969, 0.969}\color{fgcolor}
\includegraphics[width=\maxwidth]{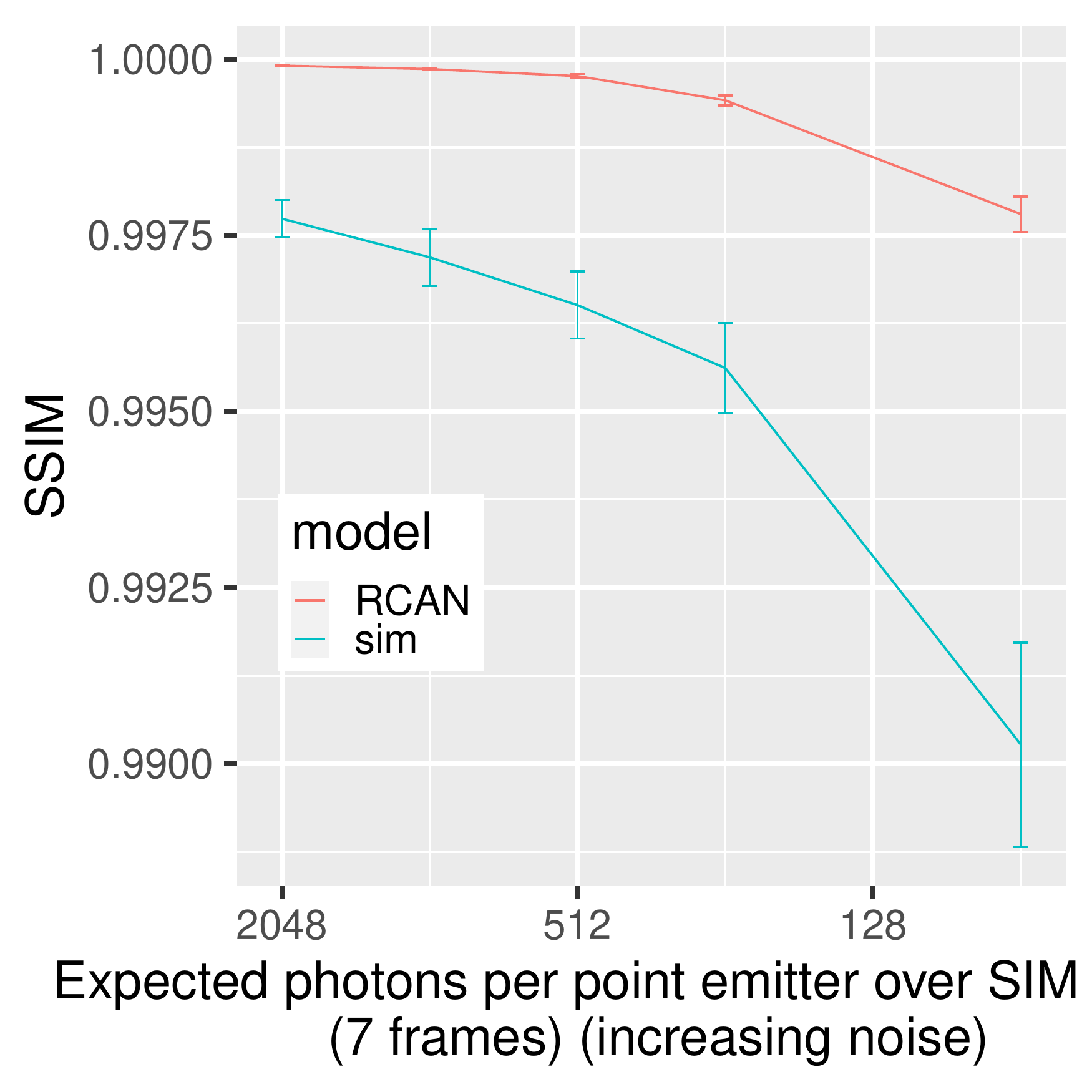} 

\end{knitrout}
    \caption{}
    \label{fig:noise_ssim}
\end{subfigure}
\caption{The effect of Poisson noise on quality of image reconstruction, run on 5 different simulated chromatin structures with 9 levels of Poisson noise. Error bars illustrate the standard error. Dotted lines in \ref{fig:noise_z} and \ref{fig:noise_xy} illustrate the theoretical resolution for the wide-field images (black) and SIM reconstructions (blue). The RCAN network provides more stable axial resolutions (see figures a and b) without a major loss in image quality compared to SIM reconstructions, as shown by the divergence in MSE and SSIM (see figures c and d). The drop in resolution as noise levels increase represents a shortfall of our method for resolution estimation, as reconstructed point artifacts due to noise are not filtered when estimating the resolution. We therefore consider the measured resolution as a comparative measure of reconstruction quality rather than an  attainable resolution in experimentally noisy conditions.}
\label{fig:noise_charts}
\end{figure}

\begin{figure}
\includegraphics[width=5in]{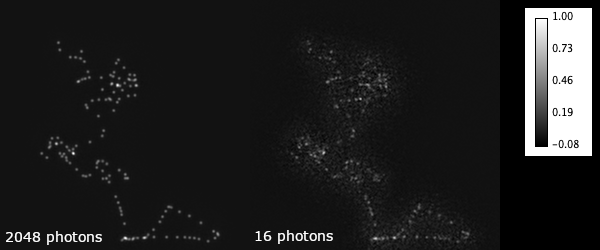}
\caption{An example of Poisson noise corrupting a sample in low-lighting conditions. The illustrated sample is chromatin  structure \#10, displayed as the axially averaged pixel value and normalised for better visibility. The left image represents a relatively well illuminated image at approximately 2048 photons per emitter across 7 SIM frames, whilst the right image illustrates an excessively attenuated lighting condition of 16 photons per emitter across 7 SIM frames. As this strength of illumination is in the order of the readout noise for CCDs, the 16 photon per emitter example serves to illustrate the form of noise implemented and was not used to evaluate the relative performance of methods in this study.}
\label{fig:rcan_noise}
\end{figure}

An additional dataset was generated using 4 other simulated chromatin structures tainted with Poisson noise. The signal-to-noise ratio was decreased by reducing the expected signal photons per emitter over 7 SIM frames, thereby simulating low-lighting conditions. The RCAN model demonstrated robustness against low-lighting conditions by maintaining a consistent axial resolution (\ref{fig:noise_z}) MSE (\ref{fig:noise_mse}) and SSIM \cite{ssim} (\ref{fig:noise_ssim}), in contrast to the reconstructed SIM images which diverge from the reference HR images in SSIM and MSE in worsening lighting conditions. The notable decrease in axial resolution for SIM reconstructions at lower lighting levels is due to a breakdown in the quality of the reconstruction, as reflected by the large increase in MSE and decrease in SSIM, whilst these are more consistent for RCAN reconstructions.

\subsection{Evaluation against third party data}
\begin{figure}[H]
\centering
\includegraphics[width=0.5\textwidth]{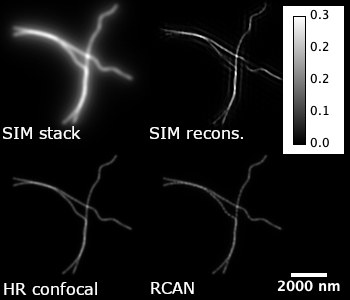}
\caption{Axial mean of the MT1.N1.HD SMLM dataset \cite{epfl_smlm_comp}; simulated SIM stack (top-left), SIM-reconstructed stack (top-right), high-res confocal image (bottom-left) and RCAN output (bottom-right). Note the high structural fidelity and sharpness of the RCAN reconstruction, without the presence of faint side-lobe artifacts seen in the SIM reconstruction. These are visible as duplicate chromatin microtubules shifted on either side of the well illuminated structure, and are more apparent in the full video of the image stacks (available in the supplementary material). All images are rescaled from 8-bit pixel values to a range of [0,1] unless they were already outputted in that format, and no normalisation was done.}
\label{fig:external_data}
\end{figure}

\begin{figure}[H]
\centering
\includegraphics[width=0.75\textwidth]{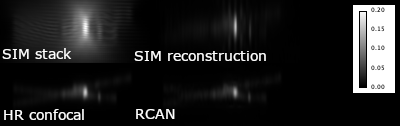}
    \caption{X/Z view of the MT1.N1.HD SMLM dataset \cite{epfl_smlm_comp}, averaged over Y.}
\label{fig:external_data_xz}
\end{figure}

The training data examined above differs from the biological structures observed in other SIM experiments, such as actin filaments \cite{unet_sim}, membranes \cite{christensen2020mlsim} or other biological structures which form wider continuous illuminations. To evaluate our network's performance on data containing these structures, cloud point data for Single Molecule Localisation Microscopy (SMLM) was taken from a SMLM competition's training dataset \cite{epfl_smlm_comp} (training sets MT1.N1.LD ($\approx$ 8 000 points, low-density) and MT1.N1.HD ($\approx$13 000 points, high-density)). These contain shallow and narrow filament structures with high point densities, allowing us to evaluate the generalisability of our method to biological structures observed in other studies. SIM image stacks and high-resolution confocal images were simulated from this data, and processed by the RCAN network. As visualised in \ref{fig:external_data}, the RCAN (bottom-right) was able to produce the expected microtubule without the side-lobe artifacts seen in the SIM image; these are visible as faint duplicate microtubules on either side of a well illuminated strand, and are more distinguishable to non-expert viewers through the video provided in the supplementary material. The difference in axial resolution (RCAN mean: 331.4nm, SIM mean: 605.1nm, reference HR confocal mean: 372.9nm) reflected the previous test on chromatin data, as did the similarity in lateral resolution (RCAN: 170.8nm, SIM: 165.7nm, reference: 172.8nm), mse (RCAN: 1.6e-05, SIM: 4.4e-05) and SSIM (RCAN: 0.998, SIM: 0.996). {A notable side-effect of the RCAN is the reconstruction of dense signal into a series of discrete point sources; this is visible in \ref{fig:external_data} when comparing the RCAN reconstruction to the HR confocal, and in Supplementary materials A7 - Figure 10b, where a uniformly dense grating is reconstructed inconsistently. This effect is likely due to the use of sparse point clouds as training data, and could be remedied in future work by simulating artificial specimens containing uniformly dense light sources.

\subsection{Execution time}
The execution time of our RCAN model was measured by computing SIM reconstructions of large image stacks. Reconstruction of 24 image stacks (280 frames each, i.e 40 SIM images with 7 frames per image) was repeated 10 times, allowing us to mitigate any file system interference due to shared use of the compute server. PyTorch's \emph{DataParallel} module enables the distribution of image chunks from the same image stack to an array GPU computation units, thereby further accelerating the reconstruction of a 3D image stack. A mean execution time of 0.88 seconds per image stack (24ms per 7-frame SIM reconstruction) was demonstrated using a shared GPU cluster of two Nvidia GeForce RTX 2080 Ti. Further GPU parallelisation may be worthwhile for larger image stacks though I/O overheads allow no further performance gains for our test data.

\section{Discussion}
\label{sec:discussion}
As demonstrated by the results in \Cref{sec:results}, the RCAN architecture described in \Cref{sec:methods} is capable of producing SIM reconstructions with a higher axial resolution than current 2D SIM reconstructions without a loss of lateral resolution or structural fidelity. This was demonstrated to be robust in low-light levels, which is typically a complexity of SIM and had been previously addressed by other deep-learning solutions \cite{unet_sim}. \newline

The use of two simulated structure types in our training data (biological structures and sphere-bounded point-clouds) demonstrates some capacity for the network to generalise to varying structures in observed specimens; this was reflected by the performance of the network on simulated SMLM data in \ref{fig:external_data}. The network was found to approximate reconstructions of continuous objects (microtubules) by reproducing the point-source patterns found in our training data. The task of generalising to other sources of data was addressed in varying approaches, such as training with real-world images \cite{christensen2020mlsim}. This mitigates the likelihood of over-fitting a model to a distinct biological structure, as they are not present across all training data. In contrast, a U-Net approach found the need to re-train models to mitigate the occurrence of artifacts in reconstructed images \cite{unet_sim}. Therefore it may be necessary to train on a wider variety of real or synthetic datasets to optimise the reconstruction of other biological structures using this method. \newline

An issue which has yet to be addressed by image up-scaling networks is the limits of the up-scaling factor with regards to image fidelity. The original implementation of the RCAN architecture \cite{rcan_zhang} demonstrates $4\times$ up-scaling but with a severe degradation of SSIM; this brings into question the certainty with which reconstructed images can be considered, especially if the ratio of up-scaling is further increased. The modified RCAN described here may be more effective when attempting higher ratios of up-scaling as the initial set of information contains many more frames (i.e channels) per output image, and therefore has the capacity to contain more information to be reconstructed into the HR image. A brief insight into the effect of removing one of these channels (see supplementary material A3) demonstrates the use of all channels in the reconstruction of the image, as the removal of any single image deteriorates the MSE and SSIM statistics in comparison to the reconstruction without masked layers. Ultimately, as the ratio of up-scaling goes beyond the capability of human validation, further imaging of structures via destructive methods (such as electron microscopy) could also validate final time points in reconstructed time lapse SIM images. A cross-validation could be performed using training groups formed of distinct biological structures; the sensitivity of the network to its training data would then provide better insight into potential reconstruction biases present in the network. This could further be extended by generating ensembles of trained models; this has previously been demonstrated in the use of image restoration for fluorescence microscopy \cite{Weigert2018}. In this scheme, each model would be trained on a random subset of the training data, and the output of all trained models for any given novel image reconstruction would be considered to form a statistical model for each output pixel. This could allow a degree of statistical confidence, which may be increasingly required to assert the certainty of reconstructions especially at larger ratios of image up-scaling.\newline

% Does image inpainting affect trustworthiness of results?
%   - Machine vs Human discussed in context of medical AI https://insightsimaging.springeropen.com/articles/10.1186/s13244-019-0785-8#Sec3
% At least we can assert frames all contribute to image

\section*{Funding}
Funding for this project is provided by
\begin{itemize}
\item the European Union (under the Horizon 2020 Framework Programme: H2020 Future and Emerging Technologies (801336 - PROCHIP)).
\item the Wellcome Trust's PhD program at Imperial College, London.
\end{itemize}

% TODO data accesssibility
\section*{Data accessibility}
\label{sec:data_access}
All code used to generate training data, define \& train deep learning models and evaluate these models is available at \url{https://github.com/mb1069/sim-axial-resolution}.

\section*{Competing interests}
The authors declare no competing interests.

\bibliographystyle{unsrt}
\bibliography{references.bib}

\end{document}

\let\ref\autoref
\title{Improving axial resolution in SIM using deep learning}
\author{Boland, Miguel; Cohen, Edward A. K.; Flaxman, Seth; Neil, Mark}
\maketitle

\section{Supplementary material A1 - Measuring FWHM}
As detailed in the main paper, the FWHM is measured from the autocorrelation function using the methodology below. The example details the estimation of the axial (Z) resolution; lateral resolution can be calculated by summing over the Z axis in \ref{1.3}, selecting the zero-padded row of matrix F in \ref{1.5} which contains the peak value for the matrix and decomposing the series into two Gaussian functions in \ref{1.6}.

\begin{equation}
\label{1.1}
\text{Let } I \text{ be a 3D microscopy X * Y * Z image stack of dimensions } (N_X,N_Y,N_Z)
\end{equation}

Let
\begin{equation}\label{1.2}
P = |FFT_3(I)|^2 \text{ where FFT3 is a 3D fast-Fourier transform.}
\end{equation}

Summing $P$ across the $X$ and $Y$ dimensions gives vector $S$ with $k$th element
\begin{equation}\label{1.3}
S_k = \sum_{i=1}^{N_X}\sum_{j=1}^{N_Y}{P_{i,j,k}}.
\end{equation}
We then pad vector $S$ using the following rule
\begin{equation}\label{1.4}
S_k = 
\begin{cases}
S_k   &\text{if } k < \frac{N_Z}{2} \\
0   &\text{if } \frac{N_Z}{2} < k < \frac{N_Z}{2} + p \\
S_{k-p} &\text{if } k > \frac{N_Z}{2} + p,
\end{cases}
\text{ where \emph{p} corresponds to the width of 0-padding.}
\end{equation}
We then inverse FFT this vector
\begin{equation}\label{1.5}
F = FFT_{shift}(IFFT(S)),
\end{equation}
and decompose it into two Gaussian functions \(g_{acf}(\alpha, \mu, \sigma)\) and \(g_{ccf}(\alpha, \mu, \sigma)\), for the auto-correlation and cross-correlation function, respectively, via a least squares fit
\begin{equation}\label{1.6}
F = g_{acf}(\alpha_{acf}, \mu_{acf}, \sigma_{acf}) + g_{ccf}(\alpha_{ccf}, \mu_{ccf}, \sigma_{ccf}).
\end{equation}
Constraints are placed that \( \mu_{acf} \approx \mu_{ccf} \) and \( \sigma_{acf} < \sigma_{ccf} \). 
The FWHM is then measured on $g_{acf}(\alpha_{acf}, \mu_{acf}, \sigma_{acf})$.

\section{Supplementary material A2 - Y/Z cross sections of output images}

\begin{figure}[H]
\centering
\begin{subfigure}{.32\textwidth}
    \centering
    \includegraphics[height=8cm,keepaspectratio]{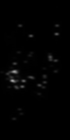}
    \caption{Reference HR}
    \label{fig:10_struct_reference}
\end{subfigure}
\begin{subfigure}{.32\textwidth}
    \centering
    \includegraphics[height=8cm,keepaspectratio]{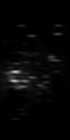}
    \caption{SIM reconstruction}
    \label{fig:10_struct_sim}
\end{subfigure}
\begin{subfigure}{.32\textwidth}
    \centering
    \includegraphics[height=8cm,keepaspectratio]{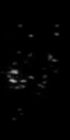}
    \caption{RCAN}
    \label{fig:10_struct_rcan}
\end{subfigure}
\caption{Y/Z cross-sections of a simulated chromatin structure}
\label{fig:10_struct_example}
\end{figure}

\section{Supplementary material A3 - Effect of obscuring a frame from the RCAN's input}
The charts below demonstrate the effect of pinning any individual SIM frame to 0 when inputting a chunk of 3 SIM reconstructions into the RCAN. This demonstrates the usage of all available input frames to reconstruct the output image, as the lowest MSE and best SSIM are only achieved when no frame is obscured.

\begin{figure}[h]
\centering
\begin{subfigure}{.45\textwidth}
\begin{knitrout}
\definecolor{shadecolor}{rgb}{0.969, 0.969, 0.969}\color{fgcolor}
\includegraphics[width=\maxwidth]{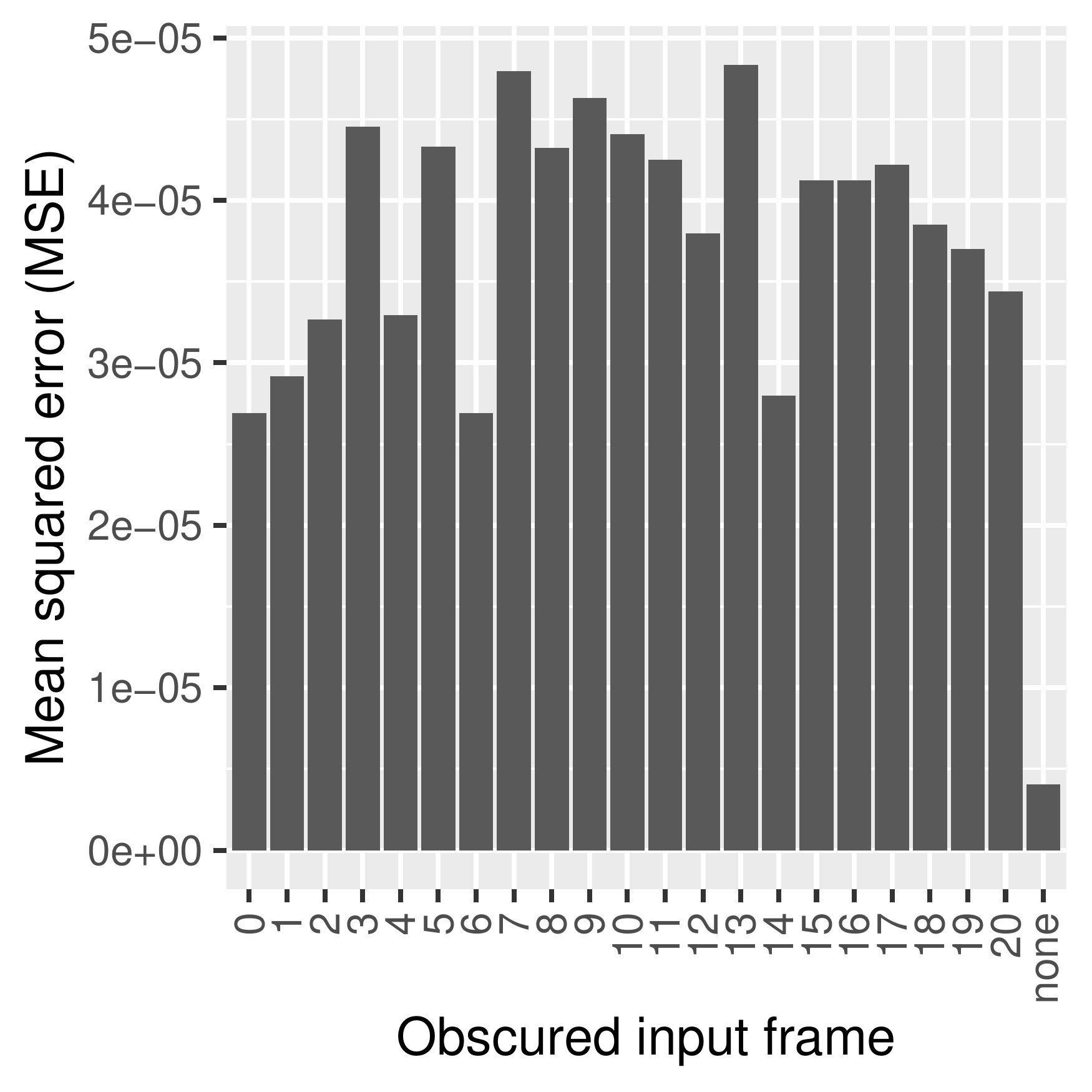} 

\end{knitrout}
\end{subfigure}
\begin{subfigure}{.45\textwidth}
\begin{knitrout}
\definecolor{shadecolor}{rgb}{0.969, 0.969, 0.969}\color{fgcolor}
\includegraphics[width=\maxwidth]{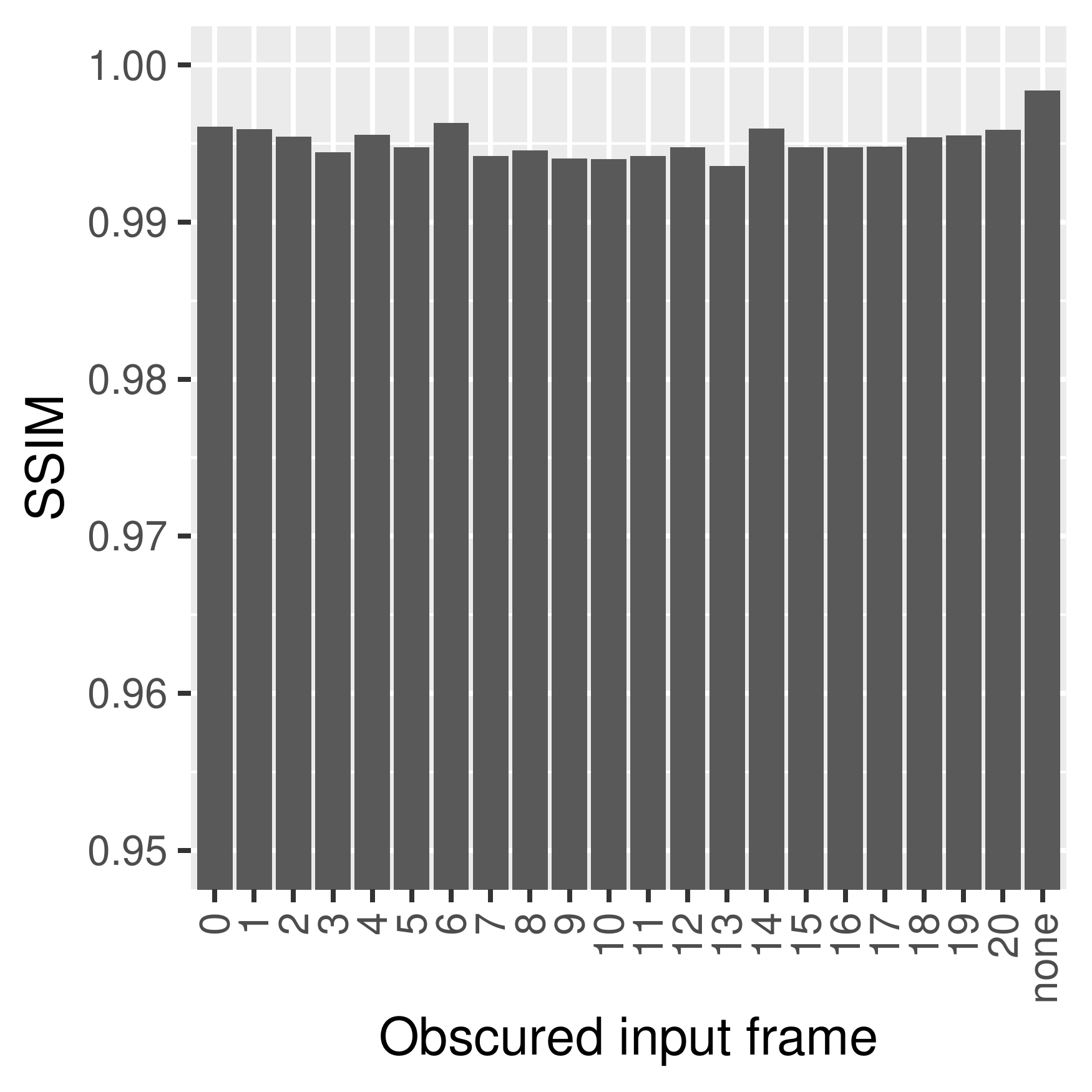} 

\end{knitrout}
\end{subfigure}
\caption{Evaluating the effect of masking one of 21 frames in an image chunk, taken from a simulated chromatin structure \& processed by our model.}
\label{fig:masked_images}
\end{figure}

\section{Supplementary material A4 - Example modelling of autocorrelation function as sum of two gaussians}
\begin{figure}[h]
\centering
    \includegraphics[width=0.75\textwidth]{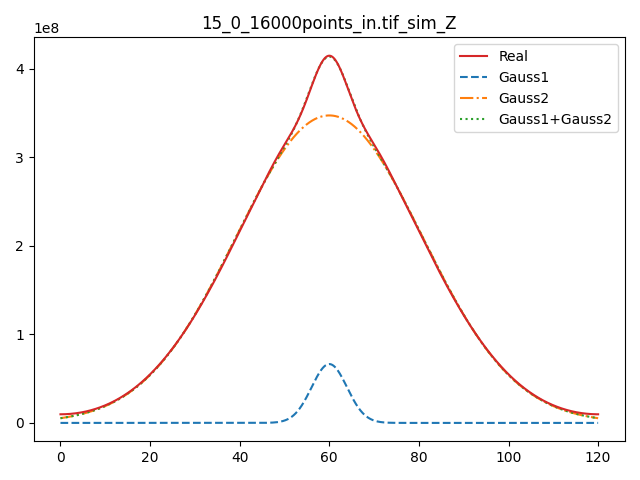}
\caption{Decomposing an image stack's inverse FFT of the power spectrum into the sum of two Gaussians}
\label{fig:gaussian_fit}
\end{figure}

\ref{fig:gaussian_fit} illustrates modelling of the inverse FFT of an image stack's power spectrum, as described in Section A1. \emph{Gauss1} is modelled as the autocorrelation function, whilst \emph{Gauss2} models the cross-correlation function; the sum of both model the measured function well as demonstrated by the \emph{Gauss1+Gauss2} series.

\section{Supplementary material A5 - Network architecture \& optimisations}
The RCAN architecture was optimised for 3D SIM reconstruction of our test datasets across a series of parameters. These are:
\begin{itemize}
    \item Number of residual groups / blocks
    \item Kernel filter size 
    \item Chunk size (number of concurrent SIM reconstructions)
\end{itemize}

As a multi-parameter grid-search optimisation would be computationally intractable, sequential optimisation was performed in the above order. Networks were trained for 50 epochs using the same dataset used in the final model's training scheme (see Methodology section of main document).

\subsection{Network depth}
\begin{figure}[H]
\centering
\begin{subfigure}{.45\textwidth}
\begin{knitrout}
\definecolor{shadecolor}{rgb}{0.969, 0.969, 0.969}\color{fgcolor}
\includegraphics[width=\maxwidth]{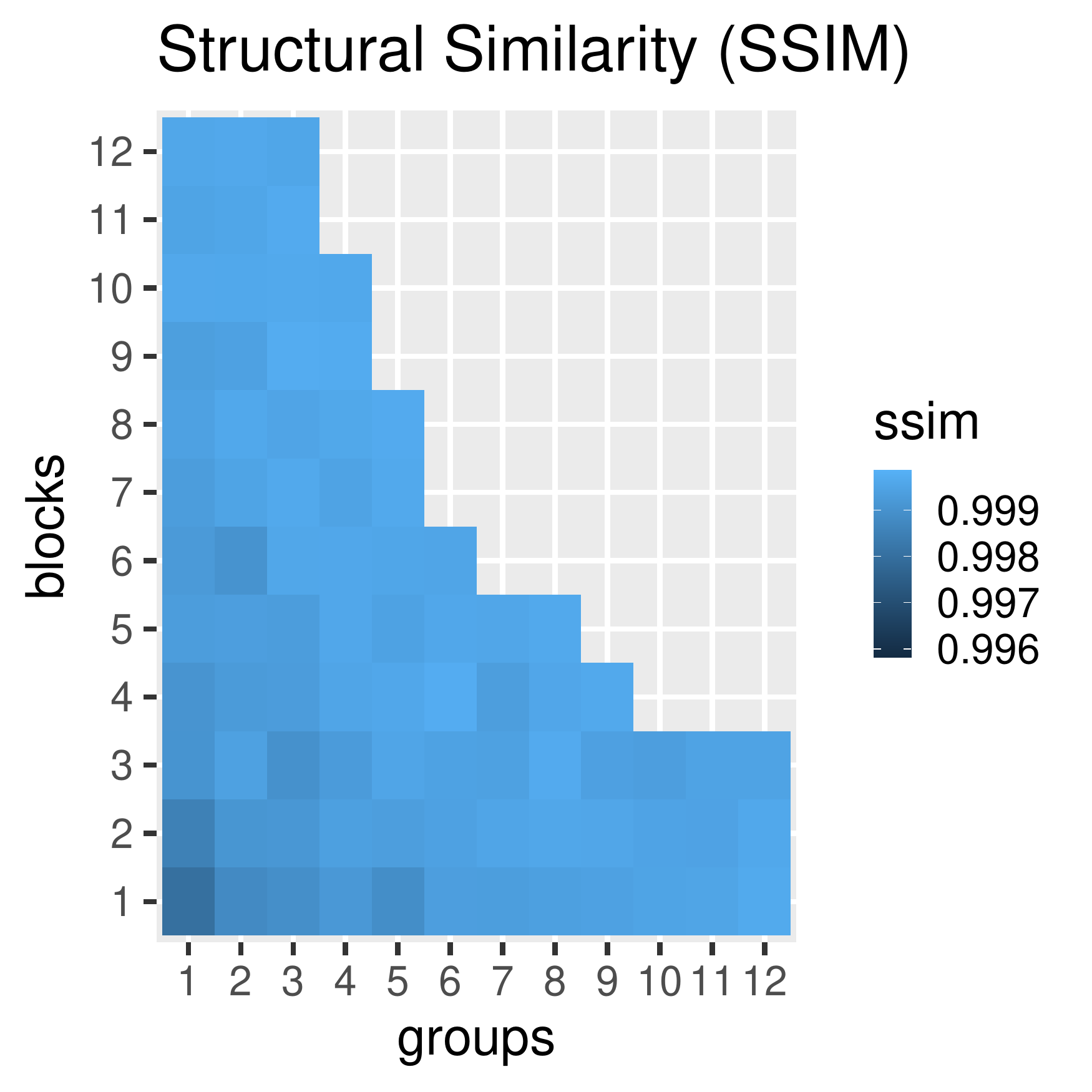} 

\end{knitrout}
\end{subfigure}
\begin{subfigure}{.45\textwidth}
\begin{knitrout}
\definecolor{shadecolor}{rgb}{0.969, 0.969, 0.969}\color{fgcolor}
\includegraphics[width=\maxwidth]{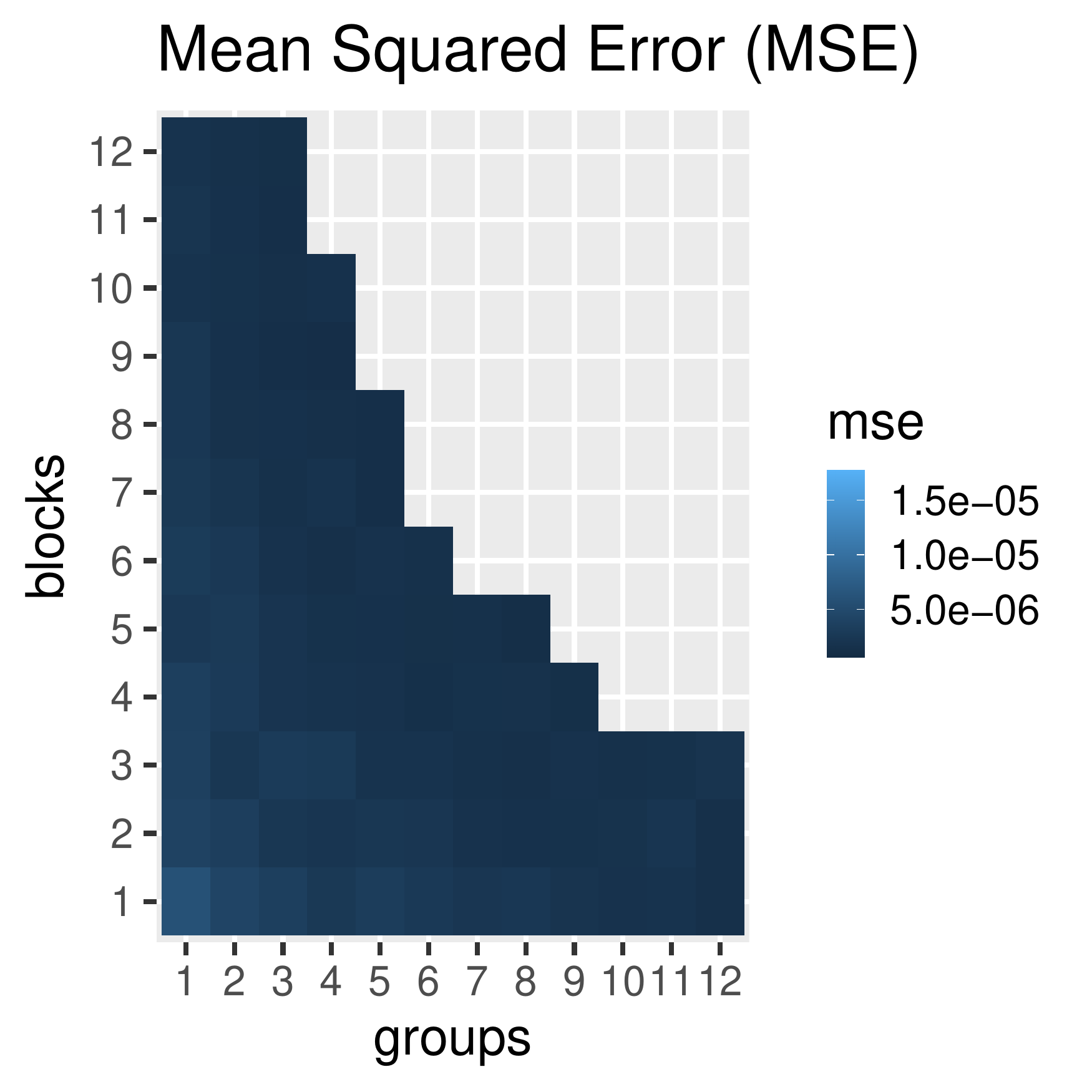} 

\end{knitrout}
\end{subfigure}
\begin{subfigure}{.45\textwidth}
\begin{knitrout}
\definecolor{shadecolor}{rgb}{0.969, 0.969, 0.969}\color{fgcolor}
\includegraphics[width=\maxwidth]{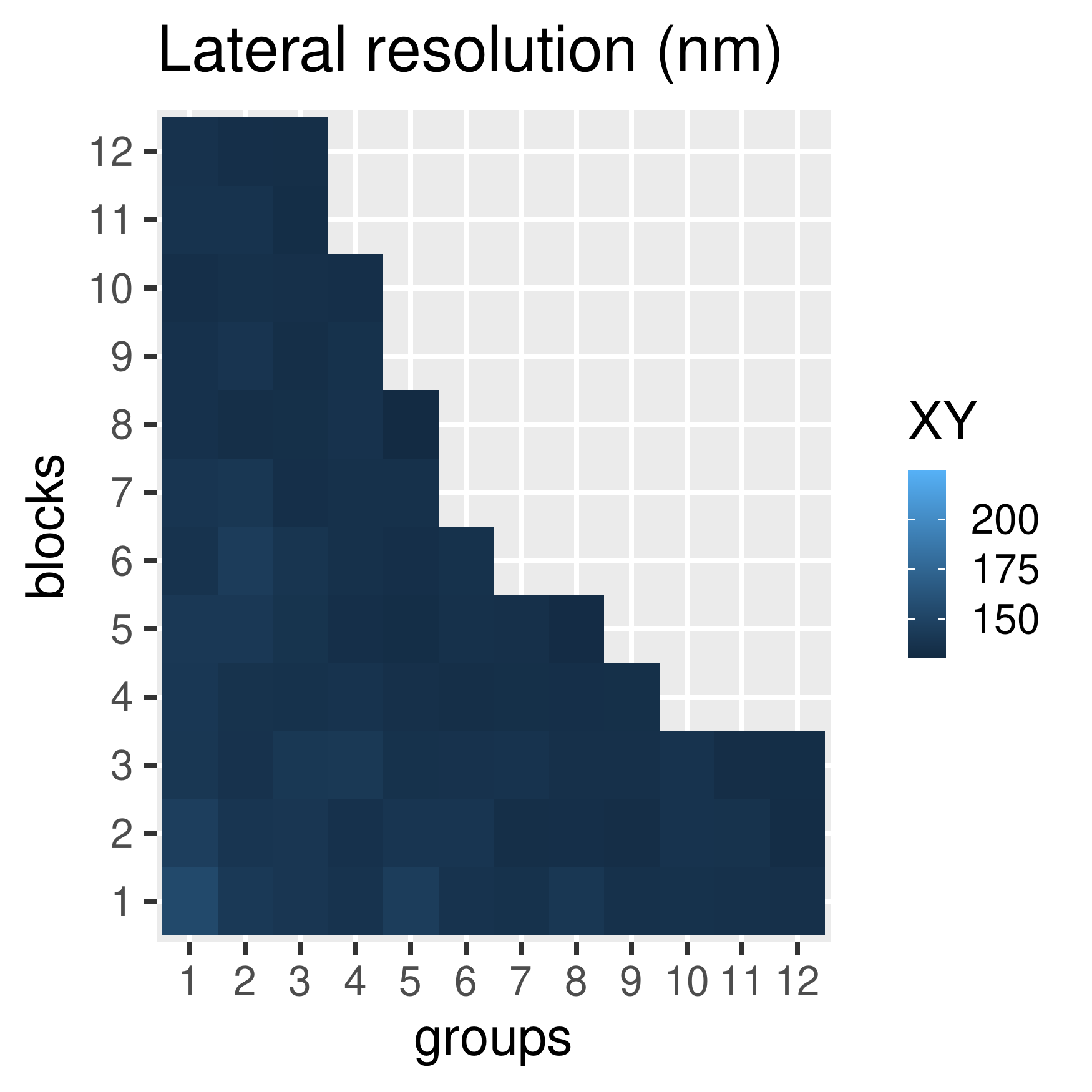} 

\end{knitrout}
\end{subfigure}
\begin{subfigure}{.45\textwidth}
\begin{knitrout}
\definecolor{shadecolor}{rgb}{0.969, 0.969, 0.969}\color{fgcolor}
\includegraphics[width=\maxwidth]{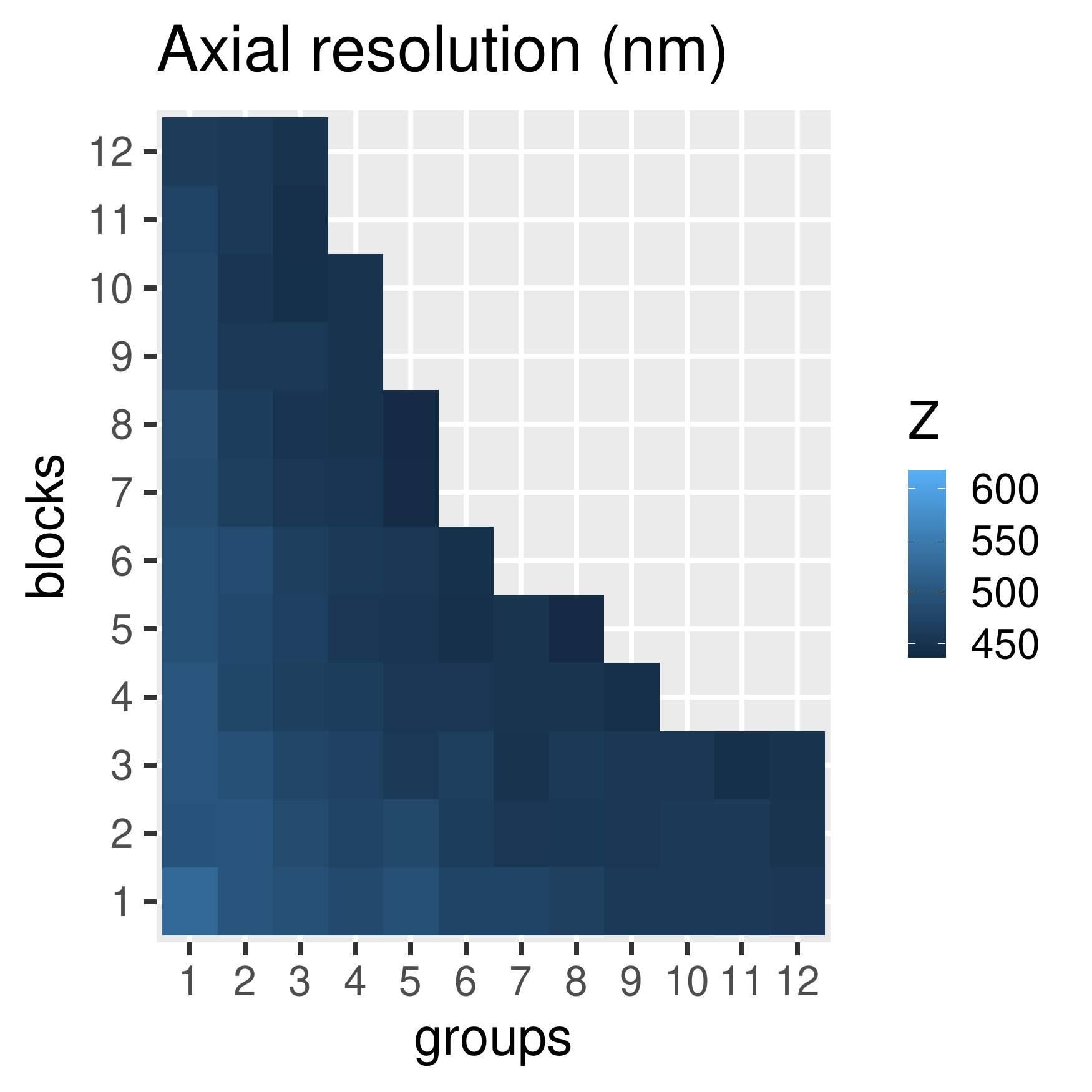} 

\end{knitrout}
\end{subfigure}
\caption{Effect of network depth and architecture on various performance metrics}
\label{fig:group_block_charts}
\end{figure}

Network depth was optimised across possible network configurations which would fit in the memory of the GPU available (Nvidia GTX 2080 TI). This ranges from networks of a single block/group up to networks of 12 groups of 3 blocks, or 3 groups of 12 blocks. Evaluation of these networks on the withheld dataset yielded the results seen in \ref{fig:group_block_charts}, and default parameters were used for the kernel size (3x3) and chunk processing (3 SIM images).

Changes in the topology of skip connections in the network (i.e change in the number of blocks and groups) has little effect on the structural similarity, MSE, or lateral and axial resolution of the network given a sufficient network depth; this is visible in the curved edge of the charts, where networks with approximately similar depth have near-indistinguishable performance. A configuration of 12 groups of 3 blocks was therefore selected due to it's marginal performance improvement across all metrics, though other similarly complex configurations may be more suitable for other training data.

\subsection{Kernel filter size}
\begin{figure}[H]
\centering
\begin{subfigure}{.45\textwidth}
\begin{knitrout}
\definecolor{shadecolor}{rgb}{0.969, 0.969, 0.969}\color{fgcolor}
\includegraphics[width=\maxwidth]{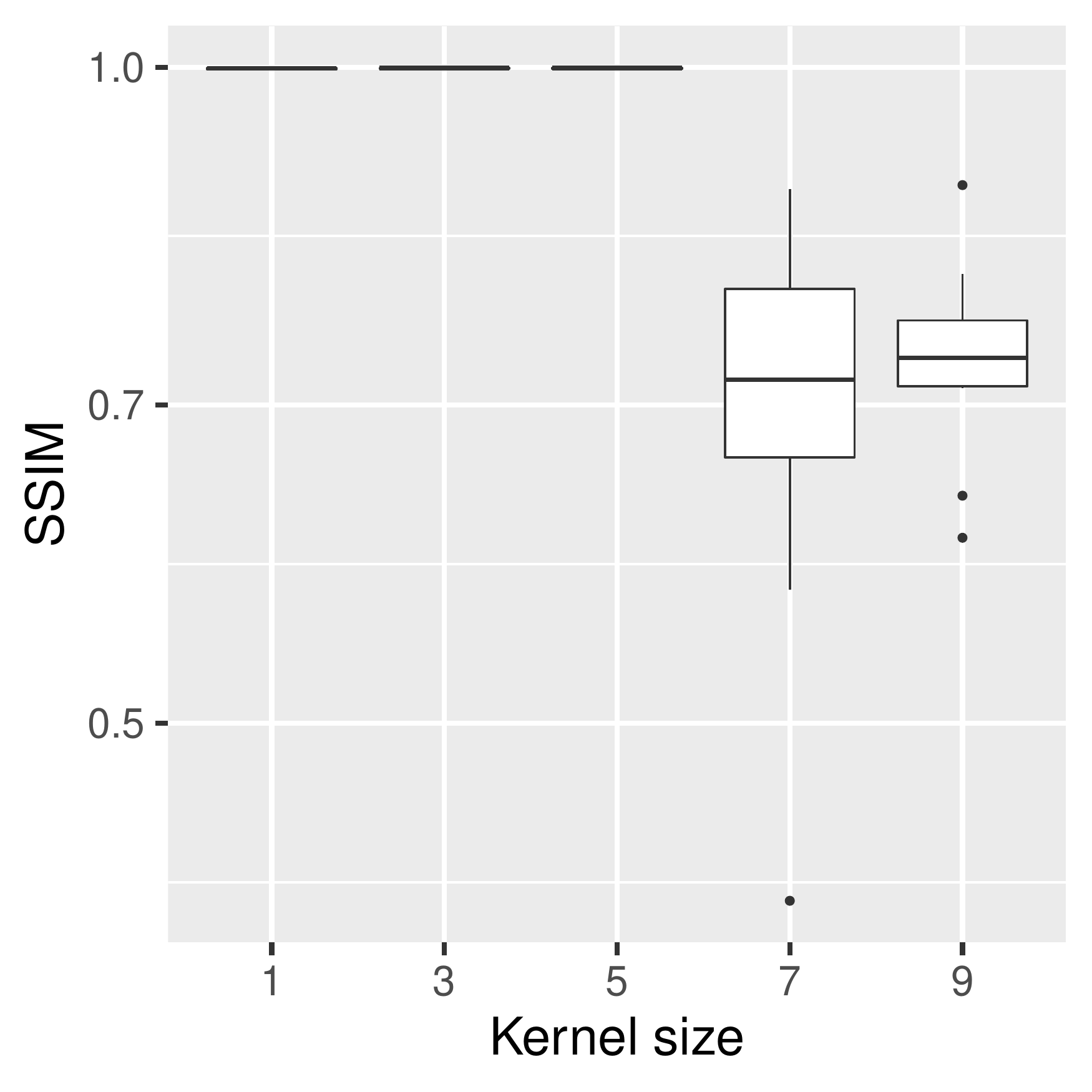} 

\end{knitrout}
\end{subfigure}
\begin{subfigure}{.45\textwidth}
\begin{knitrout}
\definecolor{shadecolor}{rgb}{0.969, 0.969, 0.969}\color{fgcolor}
\includegraphics[width=\maxwidth]{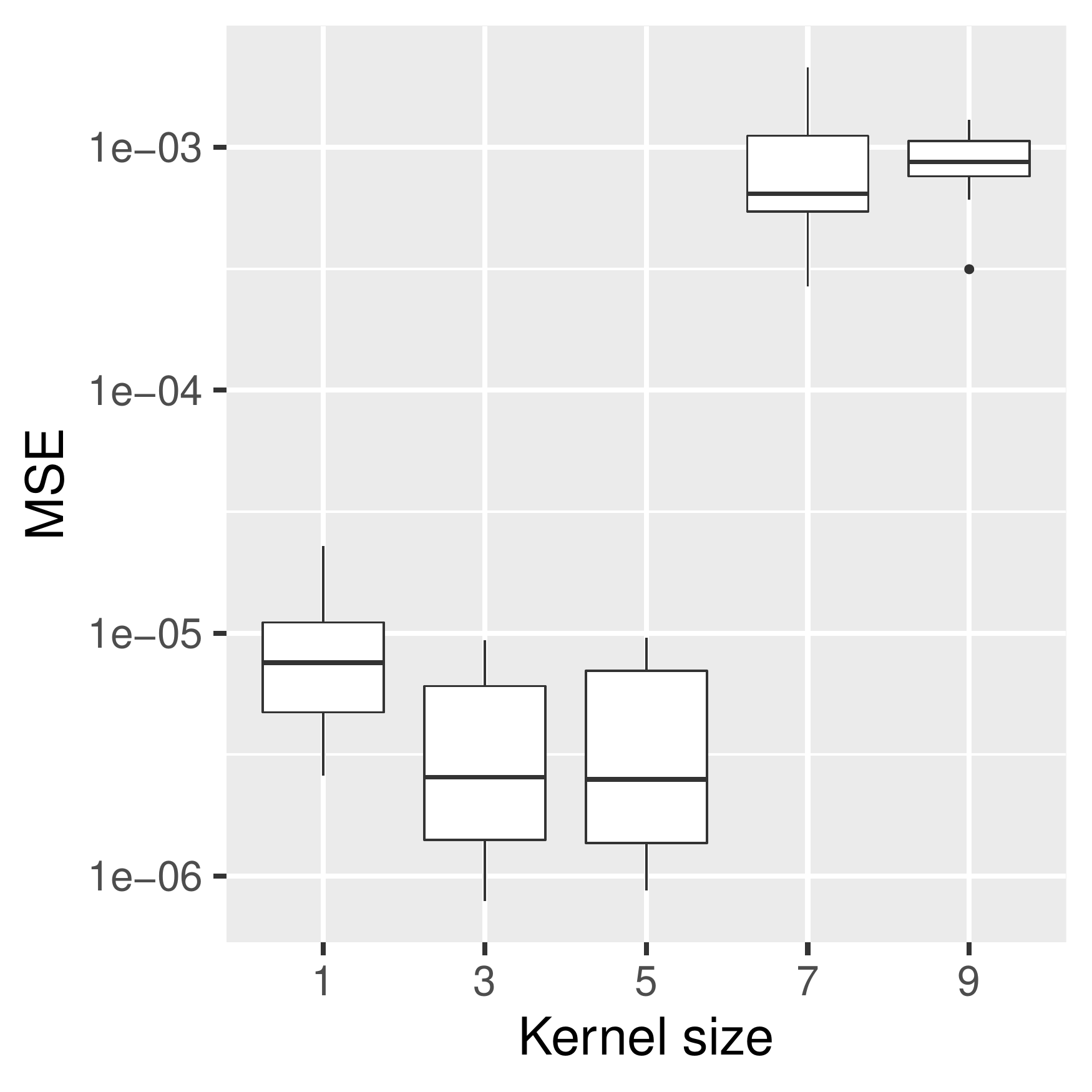} 

\end{knitrout}
\end{subfigure}
\begin{subfigure}{.45\textwidth}
\begin{knitrout}
\definecolor{shadecolor}{rgb}{0.969, 0.969, 0.969}\color{fgcolor}
\includegraphics[width=\maxwidth]{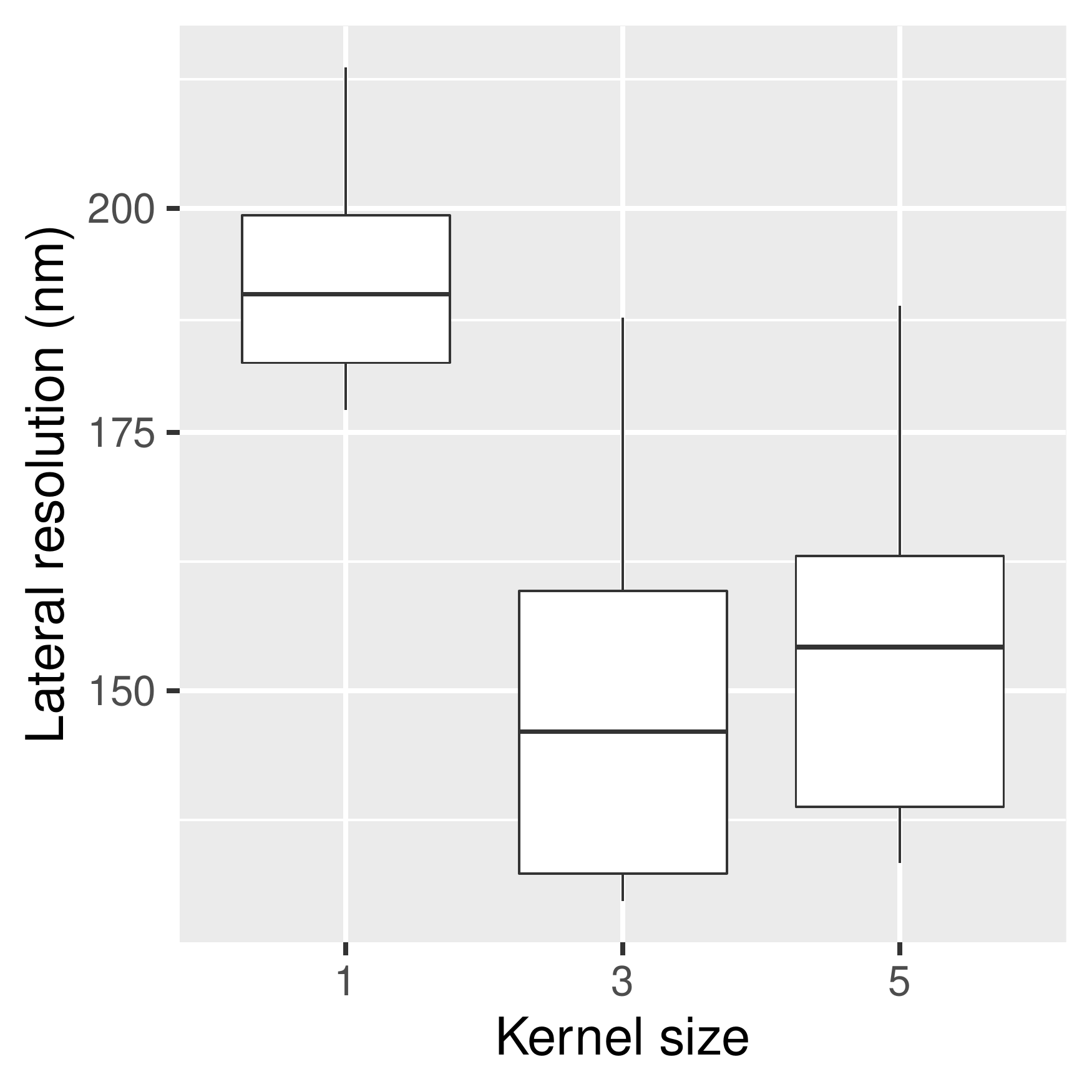} 

\end{knitrout}
\end{subfigure}
\begin{subfigure}{.45\textwidth}
\begin{knitrout}
\definecolor{shadecolor}{rgb}{0.969, 0.969, 0.969}\color{fgcolor}
\includegraphics[width=\maxwidth]{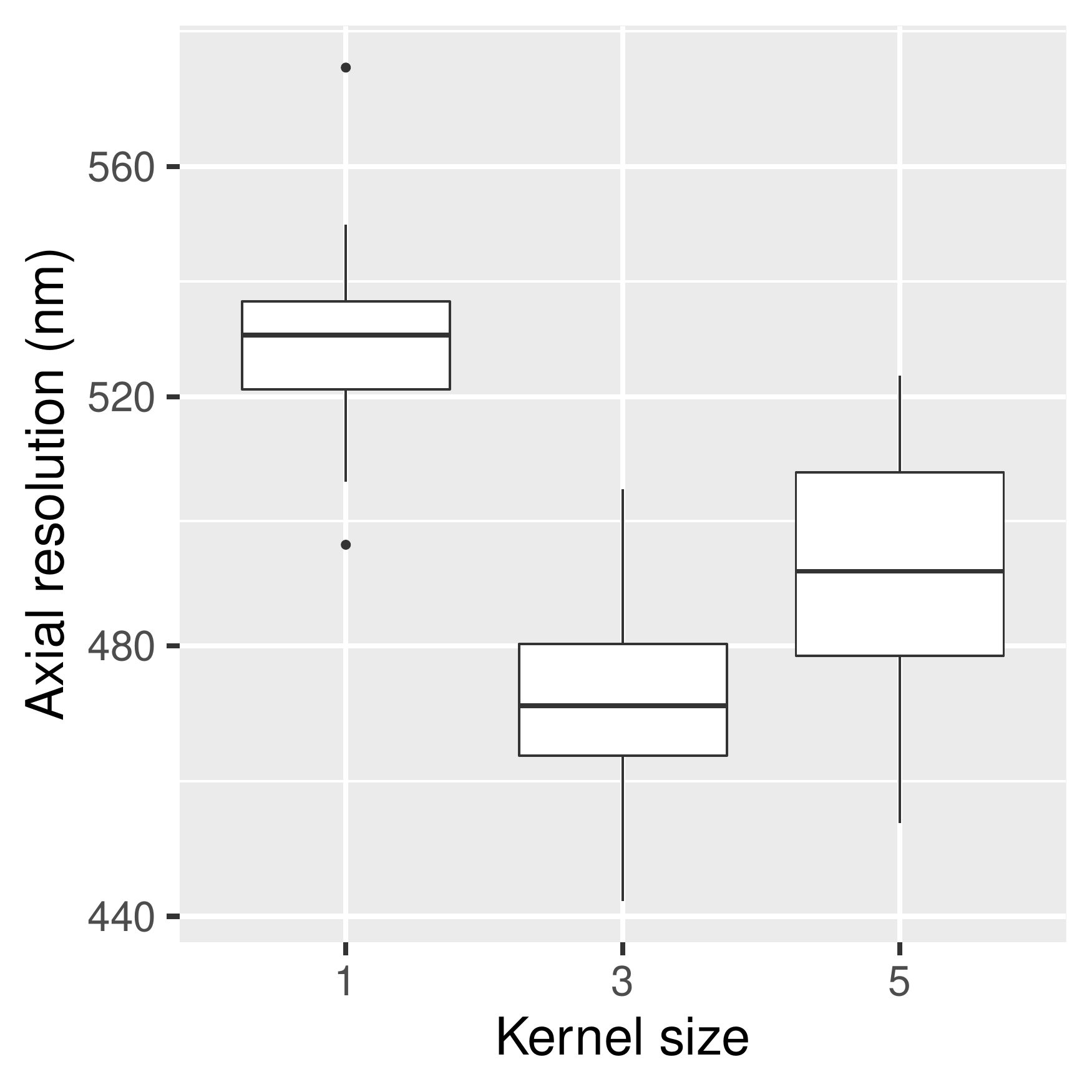} 

\end{knitrout}
\end{subfigure}
\caption{Effect on kernel filter size on network performance}
\label{fig:kernel_size_charts}
\end{figure}

Using the previously determined network architecture and a chunk size of 3, the same training scheme was applied with varying kernel filter sizes. Padding was adjusted to each kernel size so as to maintain the overall image size. As seen in \ref{fig:kernel_size_charts}, kernel sizes larger than 5 result in a complete failure of the reconstruction, demonstrated by large increases in SSIM and MSE. Of the remaining kernel sizes evaluated, a size of 3 yielded the best mean MSE, lateral resolution and axial resolution with an insignificantly different SSIM; this was therefore selected.

\subsection{Chunk size - parallel SIM reconstructions}
\begin{figure}[H]
\centering
\begin{subfigure}{.45\textwidth}
\begin{knitrout}
\definecolor{shadecolor}{rgb}{0.969, 0.969, 0.969}\color{fgcolor}
\includegraphics[width=\maxwidth]{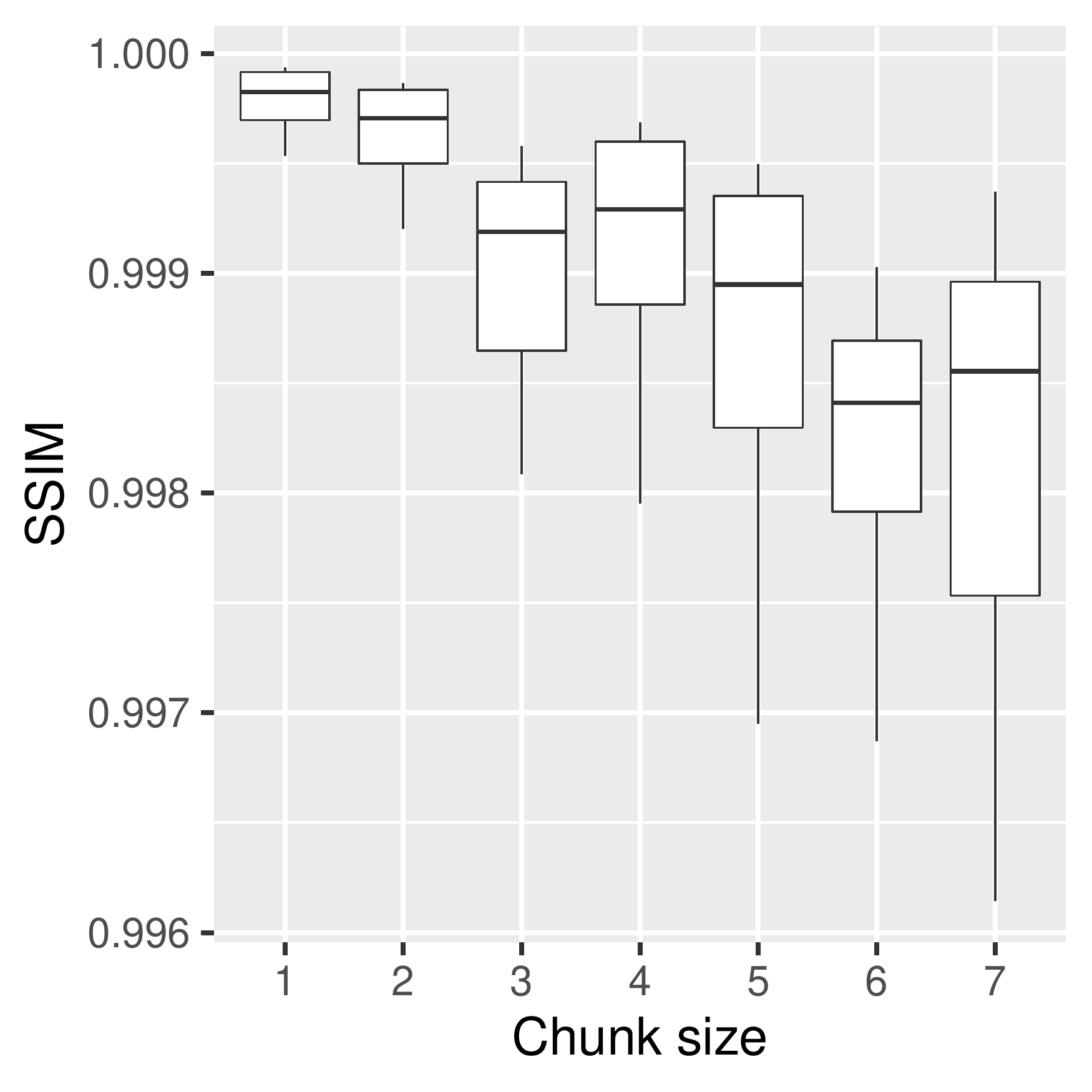} 

\end{knitrout}
\end{subfigure}
\begin{subfigure}{.45\textwidth}
\begin{knitrout}
\definecolor{shadecolor}{rgb}{0.969, 0.969, 0.969}\color{fgcolor}
\includegraphics[width=\maxwidth]{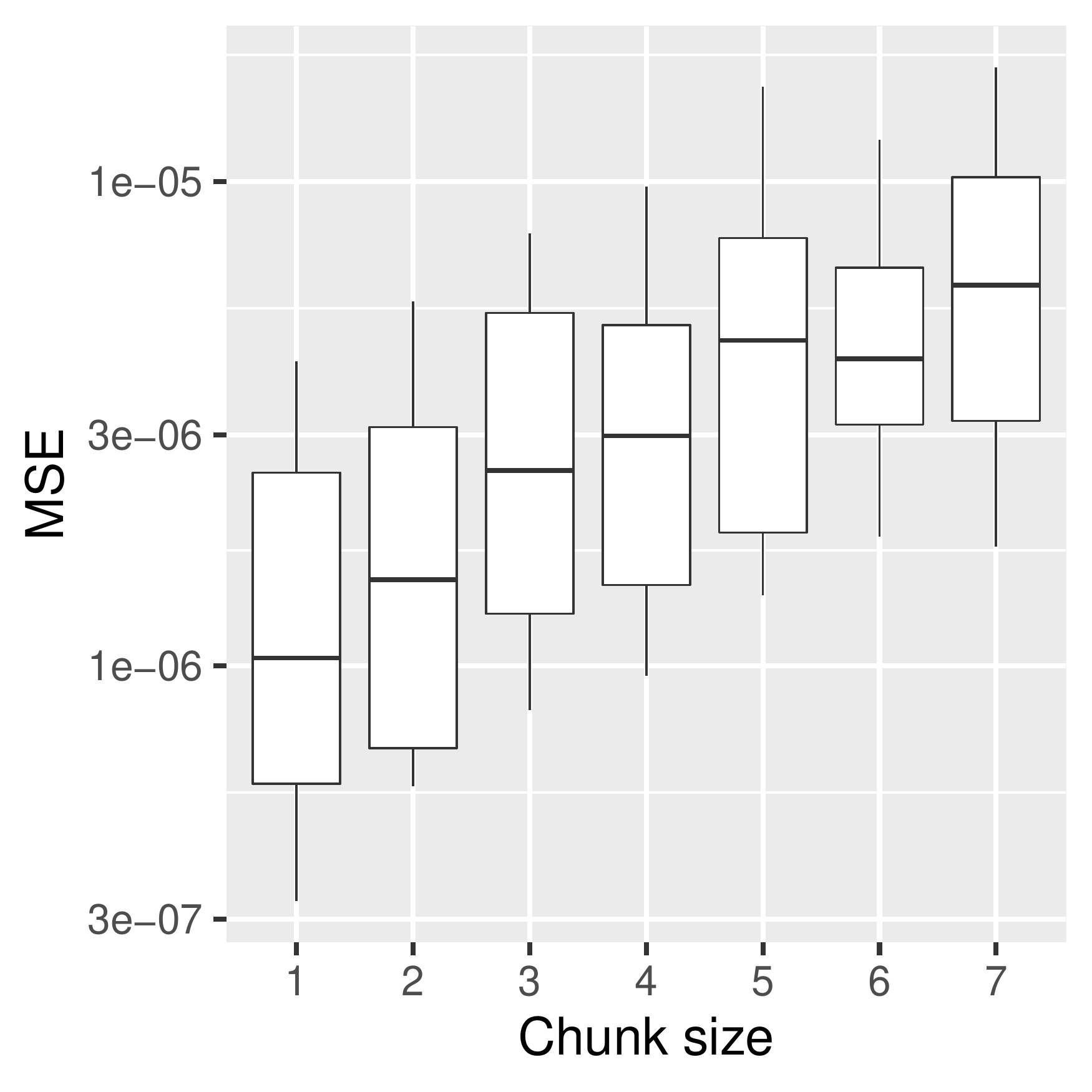} 

\end{knitrout}
\end{subfigure}
\begin{subfigure}{.45\textwidth}
\begin{knitrout}
\definecolor{shadecolor}{rgb}{0.969, 0.969, 0.969}\color{fgcolor}
\includegraphics[width=\maxwidth]{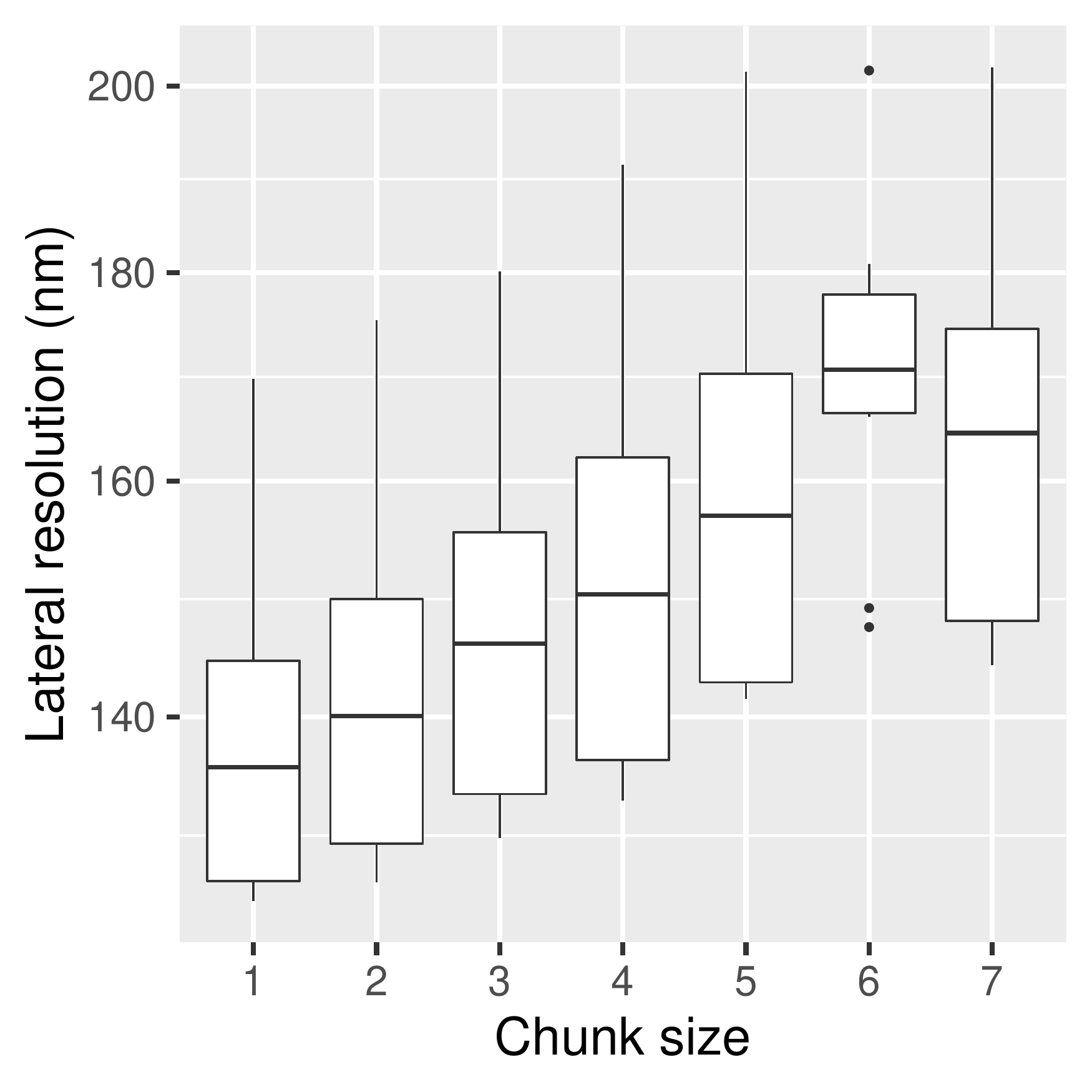} 

\end{knitrout}
\end{subfigure}
\begin{subfigure}{.45\textwidth}
\begin{knitrout}
\definecolor{shadecolor}{rgb}{0.969, 0.969, 0.969}\color{fgcolor}
\includegraphics[width=\maxwidth]{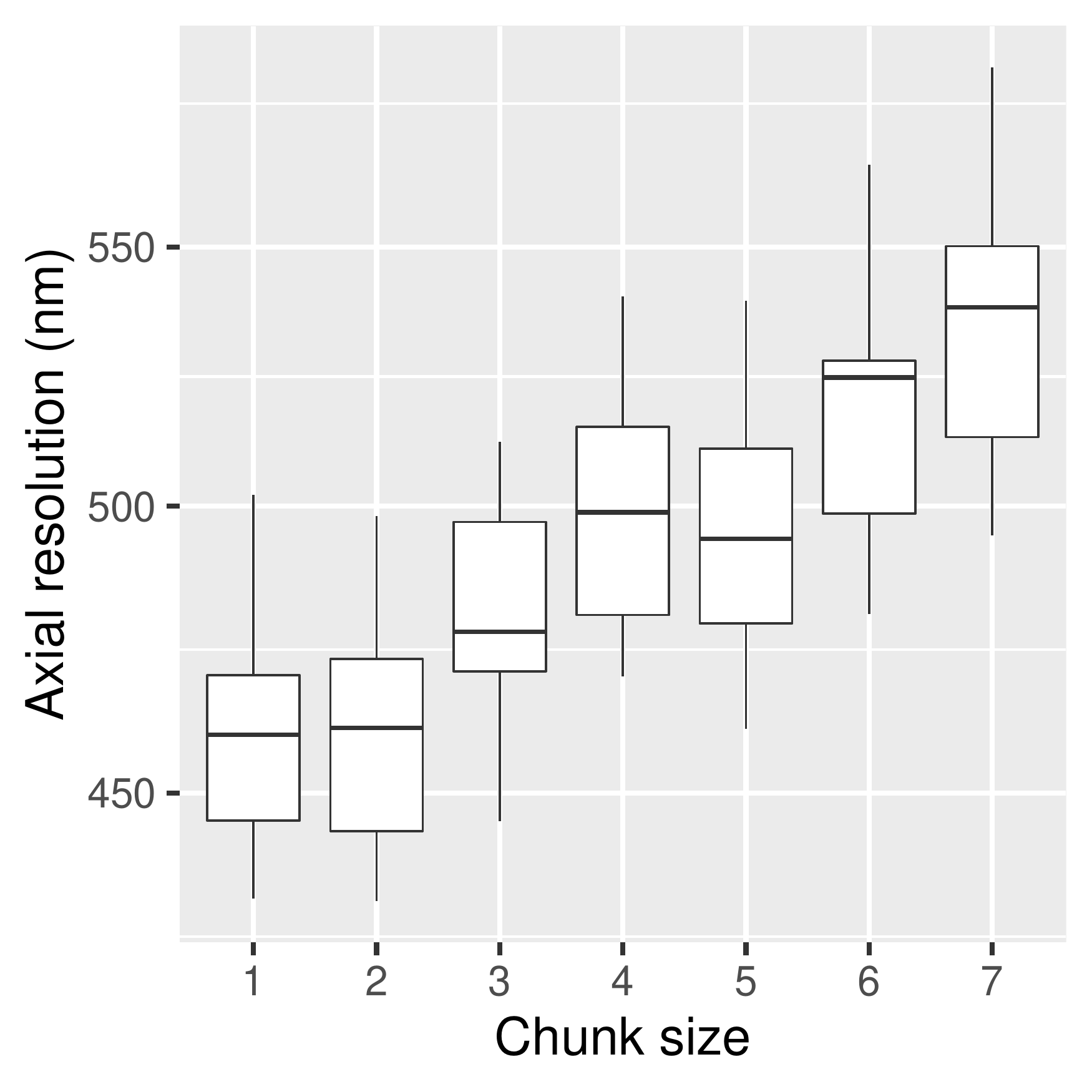} 

\end{knitrout}
\end{subfigure}
\caption{Effect of chunk size on network performance (50 epochs)}
\label{fig:chunk_size}
\end{figure}

The concurrent processing of neighbouring SIM images was assumed to allow some axial inference between reconstructions, which would be reflected in improved performance metrics. An initial evaluation (see \ref{fig:chunk_size}) over 50 epochs demonstrated a degradation of performance as we increased the number of concurrently reconstructed images; it was also noted that the network complexity increased, thereby hampering the training speed of the network. 

Repeating the experiment for 100 epochs (see \ref{fig:chunk_size_100_epochs}) demonstrated that a chunk size of 3 yielded an improvement of axial resolution (the main goal of this project) with non-significant degradation of SSIM, MSE and lateral resolution. We could further hypothesise that the degradation in performance would be minimised as the number of epochs is increased, and therefore selected a chunk size of 3 as the optimal chunk size for our dataset.

\begin{figure}[H]
\centering
\begin{subfigure}{.45\textwidth}
\begin{knitrout}
\definecolor{shadecolor}{rgb}{0.969, 0.969, 0.969}\color{fgcolor}
\includegraphics[width=\maxwidth]{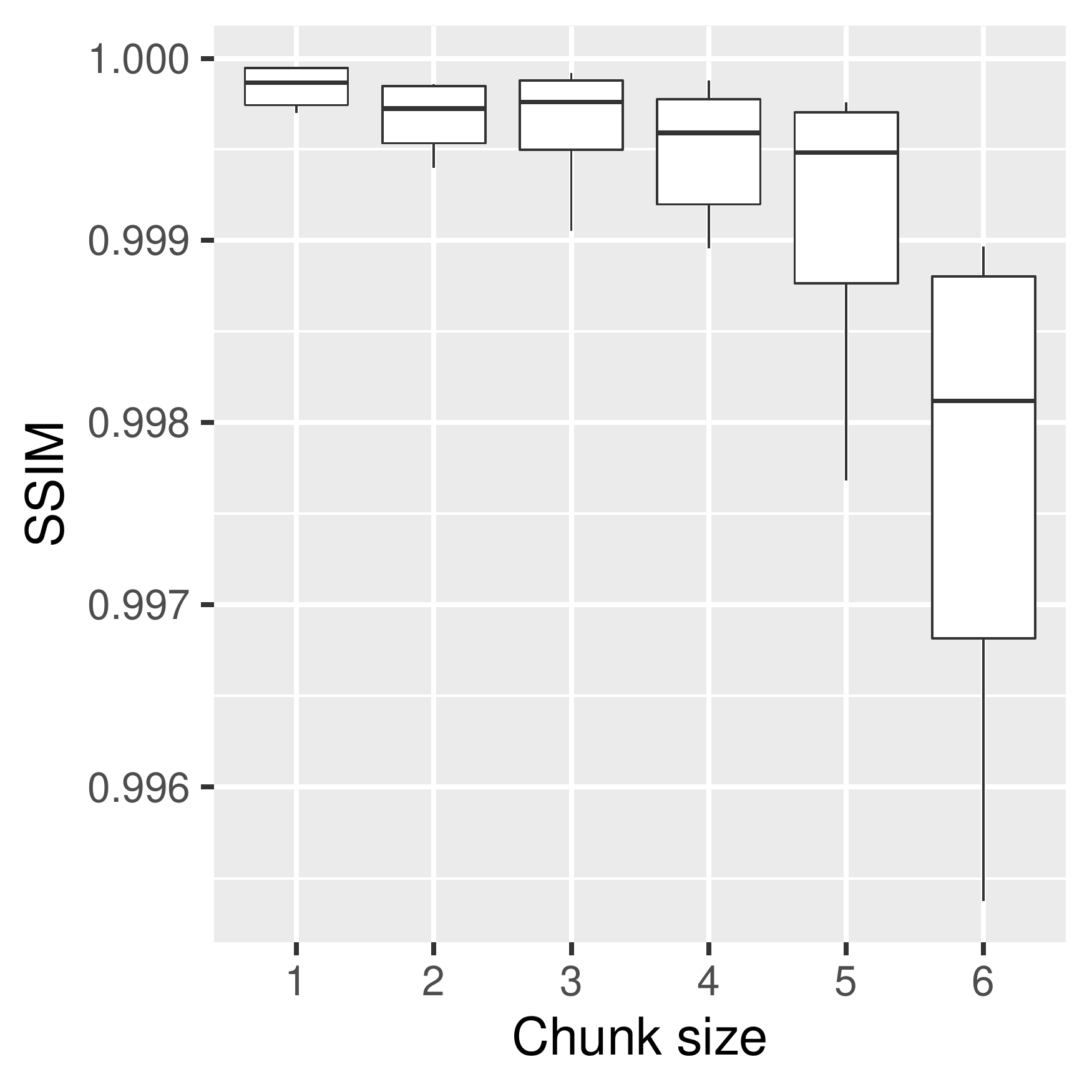} 

\end{knitrout}
\end{subfigure}
\begin{subfigure}{.45\textwidth}
\begin{knitrout}
\definecolor{shadecolor}{rgb}{0.969, 0.969, 0.969}\color{fgcolor}
\includegraphics[width=\maxwidth]{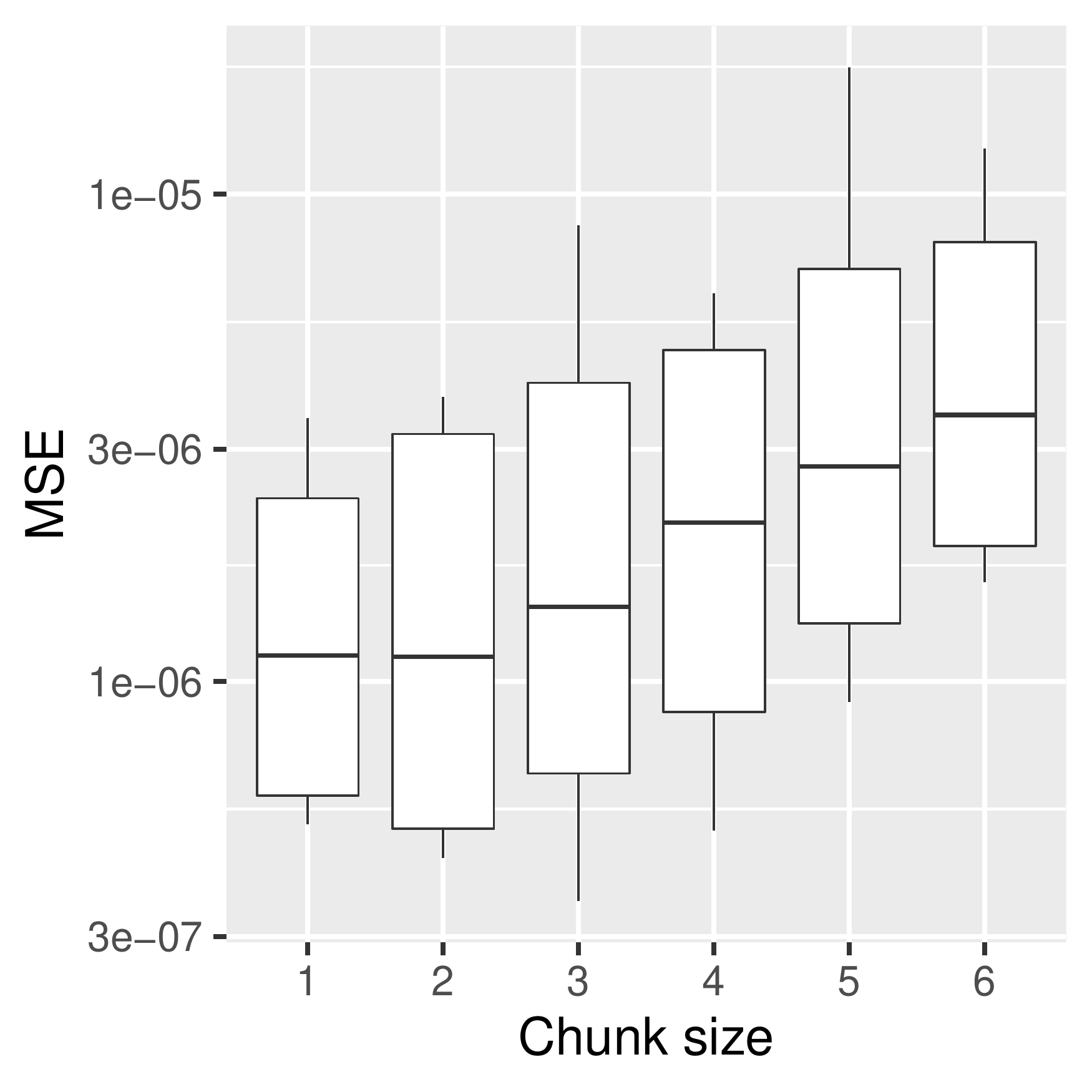} 

\end{knitrout}
\end{subfigure}
\begin{subfigure}{.45\textwidth}
\begin{knitrout}
\definecolor{shadecolor}{rgb}{0.969, 0.969, 0.969}\color{fgcolor}
\includegraphics[width=\maxwidth]{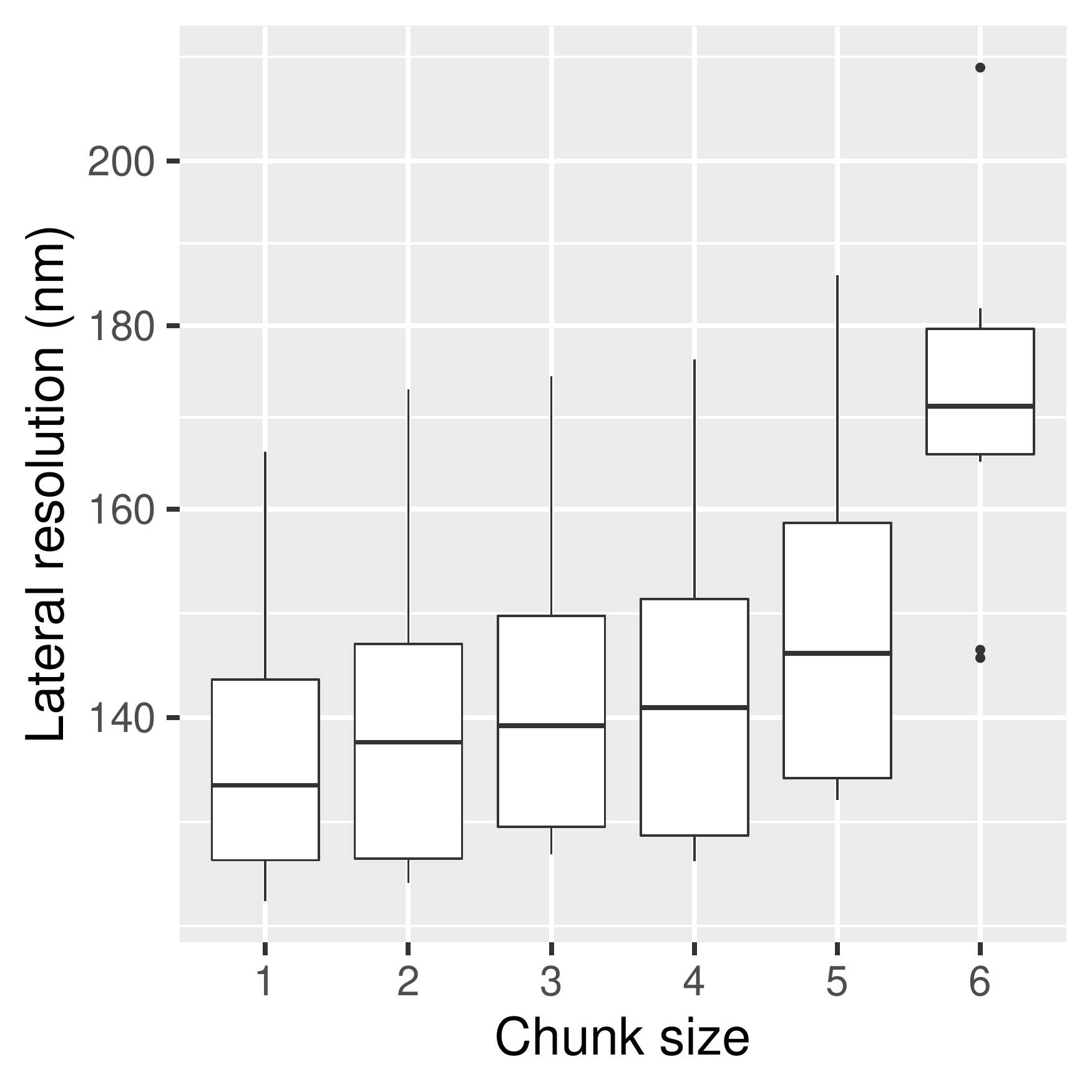} 

\end{knitrout}
\end{subfigure}
\begin{subfigure}{.45\textwidth}
\begin{knitrout}
\definecolor{shadecolor}{rgb}{0.969, 0.969, 0.969}\color{fgcolor}
\includegraphics[width=\maxwidth]{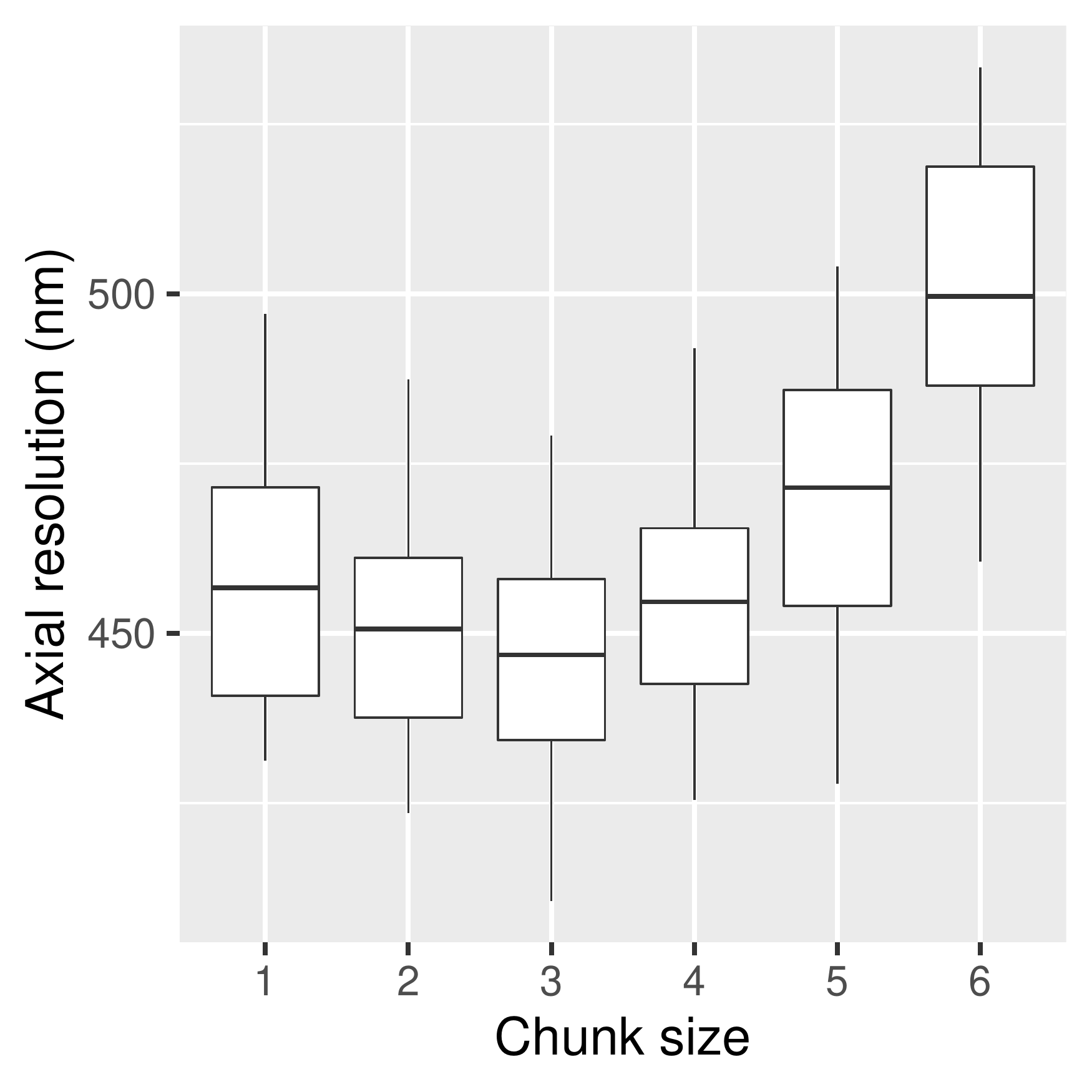} 

\end{knitrout}
\end{subfigure}
\caption{Effect of chunk size on network performance (100 epochs)}
\label{fig:chunk_size_100_epochs}
\end{figure}

\section{Supplementary material A6 - Training charts}
\begin{figure}[H]
\begin{knitrout}
\definecolor{shadecolor}{rgb}{0.969, 0.969, 0.969}\color{fgcolor}
\includegraphics[width=\maxwidth]{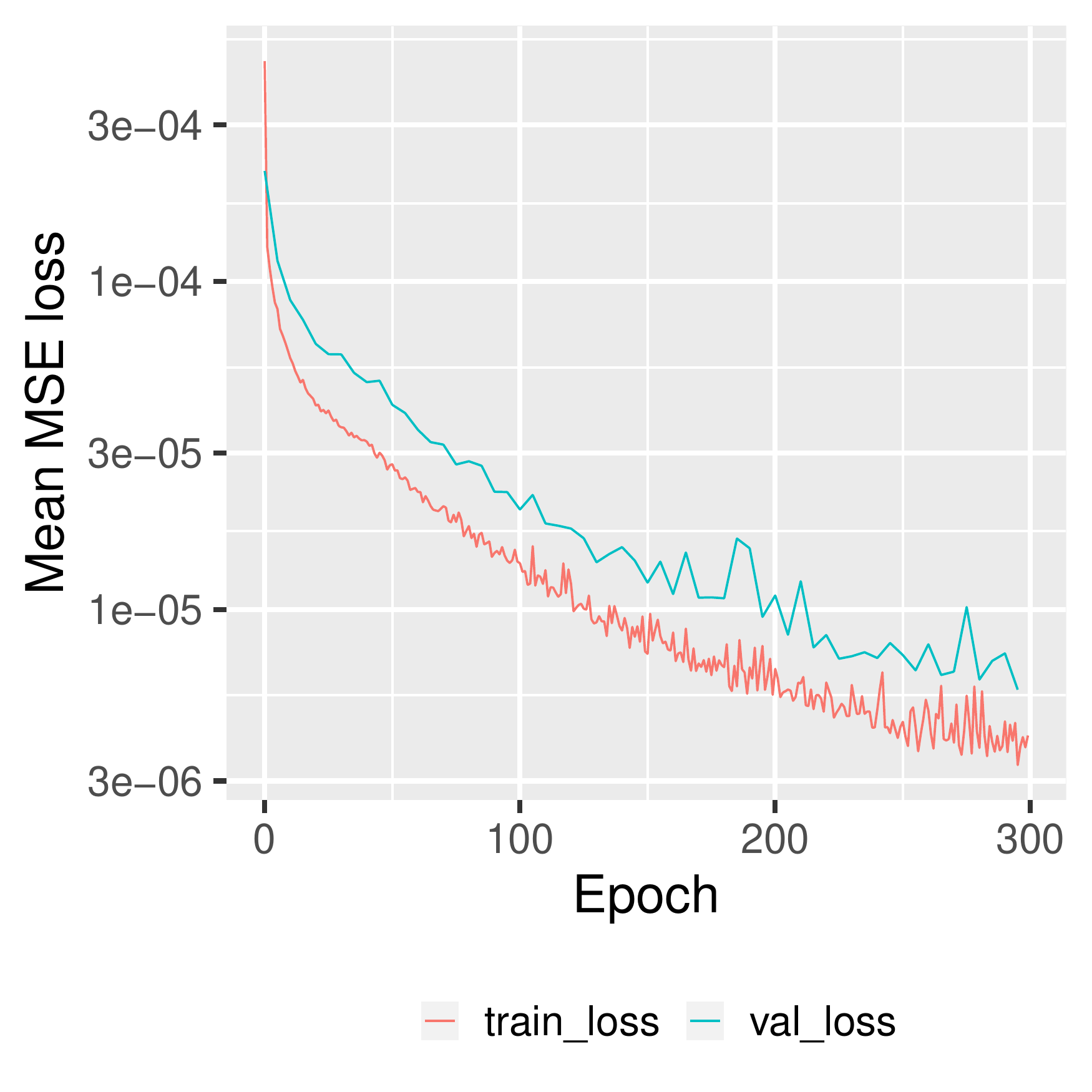} 

\end{knitrout}
\caption{An example training log, demonstrating that over-fitting is not occurring as the validation loss does not consistently increase}
\label{fig:chunk_size_100_epochs}
\end{figure}

\newpage
\section{Supplementary material A7 - Sub-resolution test cases}

The following two cases demonstrate the reconstruction of objects who's axial separation is smaller than the theoretical axial resolution of SIM but above the resolution of our RCAN model. Raw SIM image stacks were generated using the same methodology as in our main experiment; these contains either 2 points axially separated by 900nm, or 3 densely sampled planes ($1 \mu m \times 1 \mu m$, 10 000 points per $\mu m^2$) separated by 900nm. The stacks were then processed using our trained RCAN model and SIM reconstruction, the images were normalised to the range [0,1] and pixel values below 0.001 were removed to mitigate noise in the output and remove negative (imaginary) values from the images.

\begin{figure}[h]
\begin{subfigure}{0.45\textwidth}
\begin{knitrout}
\definecolor{shadecolor}{rgb}{0.969, 0.969, 0.969}\color{fgcolor}
\includegraphics[width=\maxwidth]{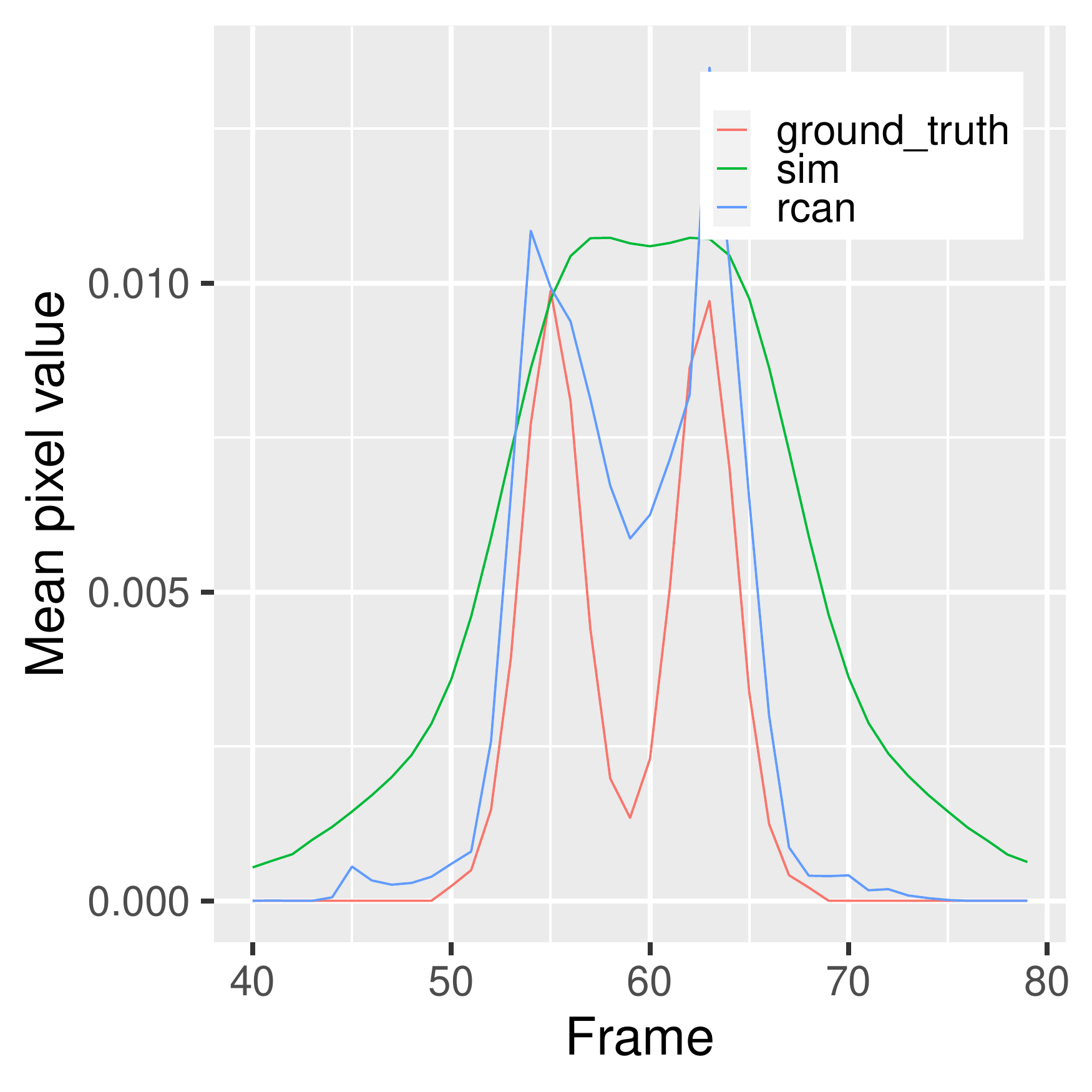} 

\end{knitrout}
\caption{2 axially separated points}
\label{fig:2point}
\end{subfigure}%
\begin{subfigure}{0.45\textwidth}
\begin{knitrout}
\definecolor{shadecolor}{rgb}{0.969, 0.969, 0.969}\color{fgcolor}
\includegraphics[width=\maxwidth]{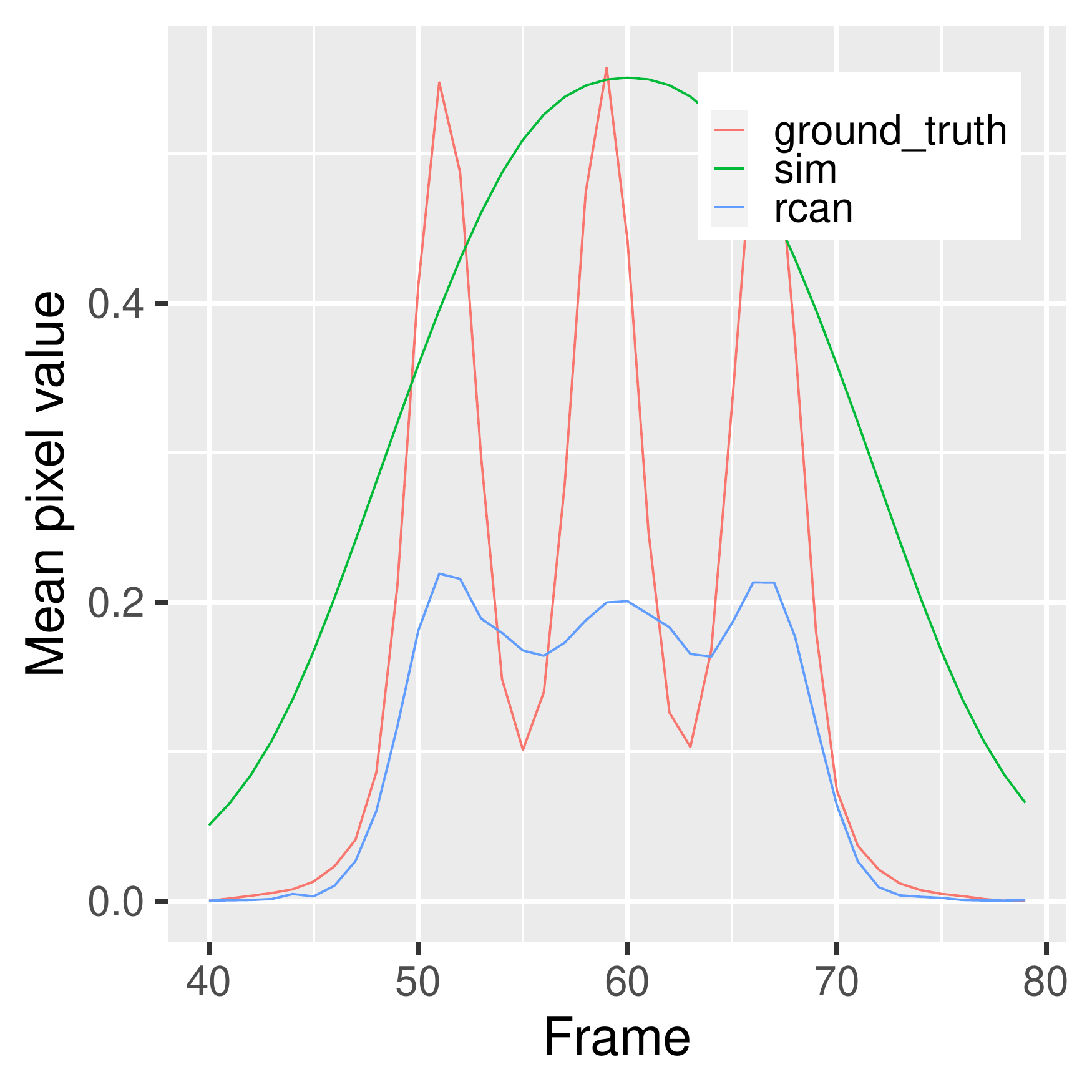} 

\end{knitrout}
\caption{Reconstruction of an axial grating}
\label{fig:grating}
\end{subfigure}
\caption{Mean pixel values along Z axis for sub-resolution test cases}
\end{figure}

As we can see in \ref{fig:2point} and \ref{fig:grating}, the SIM reconstruction fails to resolve the data into separate objects, whilst the RCAN reconstruction identifies all peaks in both cases.

\begin{figure}[h]
\begin{subfigure}{0.45\textwidth}
\begin{knitrout}
\definecolor{shadecolor}{rgb}{0.969, 0.969, 0.969}\color{fgcolor}
\includegraphics[width=\maxwidth]{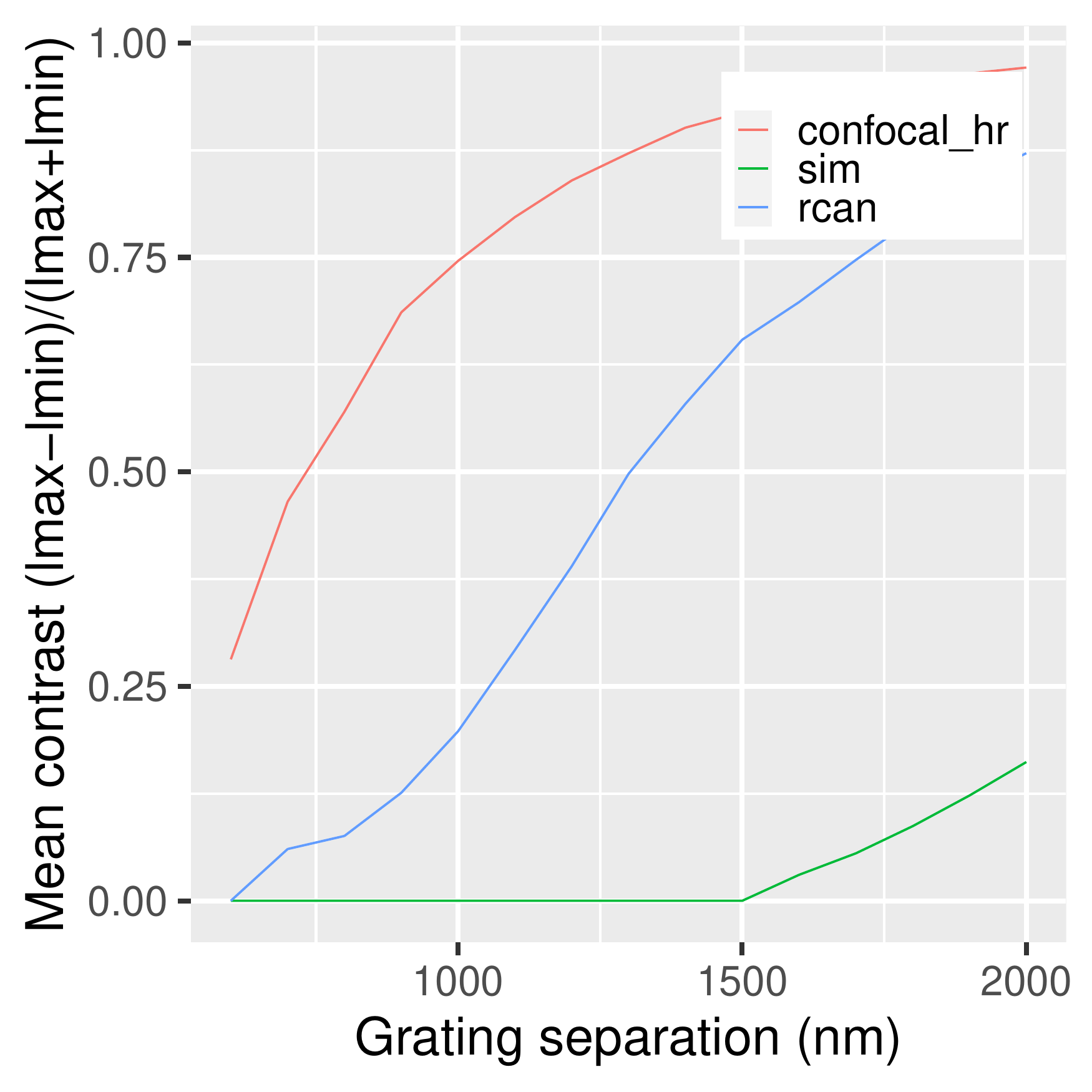} 

\end{knitrout}
\caption{Mean contrast between gratings at varying distances}
\label{fig:grating_contrast}
\end{subfigure}%
\hfill
\begin{subfigure}{0.45\textwidth}
\includegraphics[height=6cm, keepaspectratio]{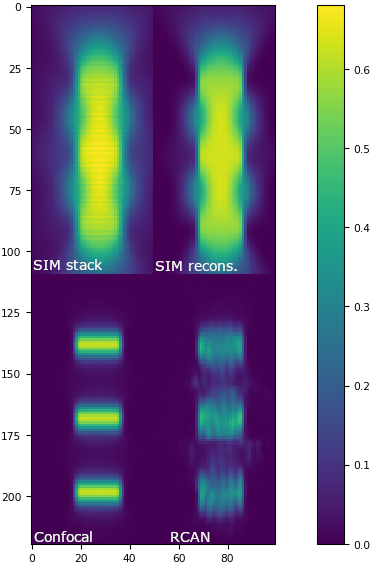}
\caption{X/Z cross-sections of 3 gratings axially separated by 1500nm.}
\label{fig:grating_contrast}
\end{subfigure}%
\caption{}
\label{fig:grating_contrast_sup}
\end{figure}
Repeating this analysis (as shown in \ref{fig:grating_contrast} over a range of grating separations demonstrates a significant improvement in contrast between sim and rcan reconstructions, though the contrast for lower resolution features could be improved in future work.